\documentclass[fleqn,usenatbib]{mnras}

\usepackage{newtxtext,newtxmath}
\usepackage{hyperref}
\usepackage{subcaption}
\usepackage{orcidlink}
\usepackage{float}
\restylefloat{table}

\usepackage[T1]{fontenc}

\DeclareRobustCommand{\VAN}[3]{#2}
\let\VANthebibliography\thebibliography
\def\thebibliography{\DeclareRobustCommand{\VAN}[3]{##3}\VANthebibliography}

\usepackage{graphicx}	% Including figure files
\usepackage{amsmath}	% Advanced maths commands
\usepackage{booktabs} 
\usepackage{multirow}
\usepackage{array}

\newcommand{\aref}[1]{\hyperref[#1]{Appendix~\ref{#1}}}

\newcommand{\vrulewidth}{0.8pt}   % thickness of vertical rule
\newcommand{\hi}{\ifmmode{\rm HI}\else{H~{\sc i}}\fi}
\newcommand{\htwo}{H\textsubscript{2}}
\newcommand{\nhi}{$N_{\mathrm{H\,\textsc{i}}}$}
\newcommand{\desp}{\textsc{despotic}}
\newcommand{\nH}{\ensuremath{n_\mathrm{H}}}
\newcommand{\anH}{\ensuremath{\langle n_\mathrm{H} \rangle}}
\newcommand{\aNH}{\ensuremath{\langle N_\mathrm{H} \rangle}}
\newcommand{\wco}{\ensuremath{W_{\rm{CO}}}}
\newcommand{\lco}{\ensuremath{L_{\rm{CO}}}}

\title[Non-equilibrium chemistry]{Modelling the non-equilibrium chemistry of the Milky Way's cold nuclear wind}

\author[Noon et al.]{Karlie A. Noon$^{\orcidlink{0000-0002-9699-6863}}$$^1$\thanks{E-mail: karlie.noon@anu.edu.au}, 
Mark R. Krumholz$^{\orcidlink{0000-0003-3893-854X}}$$^{1}$, 
Naomi M. McClure-Griffiths$^{\orcidlink{0000-0003-2730-957X}}$$^{1}$,
Enrico M. Di Teodoro$^{\orcidlink{0000-0003-4019-0673}}$$^{2,3}$
\newauthor and Lucia Armillotta$^{\orcidlink{0000-0002-5708-1927}}$$^{2,3}$
\\
$^{1}$Research School of Astronomy and Astrophysics, The Australian National University, Canberra, Australian Capital Territory, Australia\\
$^{2}$Dipartimento di Fisica e Astronomia, Università degli Studi di Firenze, via G. Sansone 1, 50019 Sesto Fiorentino, Firenze, Italy\\
$^{3}$INAF - Osservatorio Astrofisico di Arcetri, Largo E. Fermi 5, I-50125 Firenze, Italy
}

\date{Accepted XXX. Received YYY; in original form ZZZ}

\pubyear{2025}

\begin{document}
\label{firstpage}
\pagerange{\pageref{firstpage}--\pageref{lastpage}}
\maketitle

\begin{abstract}
Cold atomic and molecular gas are commonly observed in the winds of both external galaxies and the Milky Way, yet the survival and origin of these cool phases within hot galactic winds is poorly understood. To help gain insight into these problems, we carry out time-dependent chemical modelling of cool clouds in the Milky Way's nuclear wind, which possess unusual molecular-to-atomic hydrogen ratios that are inconsistent with both disc values and predictions from chemical equilibrium models. We confirm that CO and \hi{} emission comparable to that in the observed nuclear wind clouds cannot be produced by gas in chemical equilibrium, but that such conditions can be produced in a molecule-dominated cloud that has had its atomic envelope rapidly removed and has not yet reached a new chemical equilibrium. Clouds in this state harbour large reservoirs of molecular gas and consequently have anomalously large CO-to-H$_2$ conversion factors, suggesting that the masses of the observed clouds may be significantly larger than suggested by earlier analyses assuming disc-like conversions. These findings provide a new framework for interpreting cold gas in galactic winds, providing strong evidence that cold outflows can originate from the galactic disc molecular clouds that survive acceleration into the wind but lose their diffuse atomic envelopes in the process, and suggesting that the Milky Way's nuclear outflow may be more heavily mass-loaded than previously thought.

\end{abstract}


%%%%%%%%%%%%%%%%%%%%%%%%%%%%%%%%%%%%%%%%%%%%%%%%%%

%%%%%%%%%%%%%%%%% BODY OF PAPER %%%%%%%%%%%%%%%%%%

\section{Introduction}

Active star-forming galaxies, including our own Milky Way (MW), host powerful multiphase nuclear winds that shape galactic evolution. Such winds are observed from the local to the distant Universe \citep{Lehnert1995, Martin1999, Heckman2000, Pettini2001, Ajiki2002, Shapley2003, Rupke2005a}, and they play a critical role in regulating the supply of gas available for star formation and redistributing metals and energy into the circum-galactic and inter-galactic media \citep[e.g.,][]{Veilleux2005, Tumlinson2017, Zhang2018a, Veilleux2020a, Su2022}. Galactic winds are multiphase, consisting of a hot plasma at temperatures exceeding $10^6$ K \citep{Watson1984, Strickland2000}, cooler ionized \citep{Hameed1999, Veilleux2002} and neutral atomic hydrogen (\hi) components \citep{Heckman2000, Rupke2005a}, and dense, cold molecular gas  \citep{Walter2002, Bolatto2013b}. The MW wind is typical in this regard, and has been observed at wavelengths spanning the electromagnetic spectrum, including the $\gamma$-ray lobes known as the Fermi Bubbles \citep{Su2010} and the associated microwave haze likely produced by non-thermal particles (though the latter might also come from spinning dust -- \citealt{Finkbeiner2004}), the hot X-ray emitting phase at $T \sim 10^{6-7}$ K \citep[e.g.][]{Koyama1989, Snowden1997, Morris2003, Ponti2019, Predehl20}, the cool atomic hydrogen phase at $T \sim 10^{2-4}$ K \citep{Sofue1984, Lockman1984, McClure-Griffiths2013, DiTeodoro2018} and the cold molecular phase at just $T \sim 10^{1-2}$ K \citep{DiTeodoro2020, DiTeodoro2026, Cashman2021, Veena2023}. 

The coexistence of vastly different gas phases in galactic winds presents a challenge, particularly the persistence of cold molecular clouds in a surrounding hot phase. Theoretical studies of such clouds often predict their rapid destruction by hydrodynamic instabilities  \citep[e.g.,][]{Klein1994, Scannapieco2015, Schneider2017, Zhang2017}. This discrepancy between theory and observations is a fundamental problem in understanding galactic winds \citep[e.g.,][]{Zhang2017, Fielding2022}, and has led some authors to propose that the cooler material observed in galactic winds might be produced by condensation out of the hot phase rather than by survival of cool entrained material \citep[e.g.,][]{Thompson16}. On the other hand, some studies have found that clouds can persist in a hot medium under some conditions. For example, \citet{Armillotta2017} found that thermal conduction stabilises cool ($T = 10^4$ K) clouds if they are sufficiently large, while \citet{Gronke2022} found that the cold gas can survive if the cooling time of the mixed gas at the cloud-environment interface is shorter than the Kelvin–Helmholtz time. 

Given this theoretical disagreement, it is helpful to turn to observations for insight, and to do so in the MW itself, where we can achieve maximum resolution and sensitivity. The $^{12}$CO(2$\rightarrow$1) emission line is an accessible and widely used proxy for molecular gas well-suited to this goal. \cite{DiTeodoro2020} observe two molecular clouds outflowing in the MW wind selected based on their high velocities being incompatible with Galactic rotation. The clouds consist of several small clumps, seemingly embedded in diffuse \hi{} envelopes. A kinematic biconical wind model places the clouds at a distance of $0.8-1.8$ kpc from the Galactic Centre, with estimated times of $< 10$ Myr for them to travel from the Centre to their present locations \citep{Lockman2020a}. 

Using matched-resolution observations, \cite{Noon2023} find \hi{} column densities (\nhi{}) of a few $\times 10^{20}$ cm$^{-2}$ in these two clouds. This figure is highly surprising, because it implies the presence of molecular gas in regions where \nhi{} is an order of magnitude lower than is observed around essentially all Galactic molecular gas \citep[e.g.][]{Rachford2009, Lee2015}. High \hi{} columns are also expected to be required for the existence of \htwo{} from theory, because shielding \htwo{} against photodissociation by the interstellar radiation field requires a minimum column density $\approx 10^{21}$ cm$^{-2}$ at Solar metallicity \citep{Krumholz2008, Krumholz2009, McKee2010}, an order of magnitude larger than that observed. Despite this, the observed clouds maintain high molecular fractions ($\sim 50\%$).

\citet{Noon2023} propose that this anomaly can be understood if the clouds are not in chemical equilibrium. They suggest that the clouds likely originated in the disc and lost some of their \hi{} envelope while being entrained into the wind; the \htwo{} we see in them is in the process of dissociating into \hi, but has not yet had time to do so fully, explaining the anomalous \htwo{} to \hi~ratio. By contrast, if the clouds had formed within the wind through condensation or compression, they would presumably begin the process composed mostly of \hi{} before condensing or compressing further into \htwo{}. Thus the sign of the molecular disequilibrium points to a particular dynamical scenario for the origin of the cool gas.

While this hypothesis is suggestive, it has yet to be assessed quantitatively. Providing such an exploration is our principal goal here. In this paper, we investigate the chemical evolution of an atomic-molecular cloud which has experienced a stripping event (such as being swept from the disc by a wind) and whose interior is therefore now exposed to a dissociating radiation field against which it was previously shielded. We seek to ask under what circumstances such a cloud can reach a chemical state similar to what we observe in the nuclear wind clouds. The remainder of this paper is structured as follows: first, in \autoref{sec:meth}, we review the observations that we are attempting to explain, and present our methods for doing so. In \autoref{sec:equil}, we explore models of clouds in chemical equilibrium, and confirm that the observed clouds are inconsistent with being in a chemical equilibrium state. This motivates us to explore non-equilibrium models in \autoref{sec:non-equ}, where we show that it is possible to reproduce the observed chemical state of the nuclear wind clouds as resulting from non-equilibrium chemistry. 
We discuss the implications of our findings in \autoref{sec:disc} and summarise the paper in \autoref{sec:conc}.

%%%%%%%%%%%%%%%%%%%%%%%%%%
\section{Observational Constraints and Modelling Approach}
\label{sec:meth}

We begin in \autoref{ssec:observations} by reviewing the observations to which we will compare. Then in \autoref{ssec:despotic} we explain the basic structure of our chemical models.

\subsection{Observed properties of Milky Way wind clouds}
\label{ssec:observations}

We base our modelling on two clouds (C1 and C2) identified by \citet{DiTeodoro2020} as being entrained in the MW’s nuclear wind. Both clouds are detected in \hi{} and in $^{12}$CO(2$\rightarrow$1) emission. The atomic hydrogen was observed with the MeerKAT telescope, producing resolution-matched maps with the APEX CO observations \citep{Noon2023}. The resulting \nhi{} and velocity-integrated CO brightness temperature (\wco{}) maps have an angular resolution of $\sim25''$, corresponding to a physical scale of less than 1 pc at the GC. This resolution is sufficient to resolve the internal substructure within each cloud and to ensure that the two tracers probe co-spatial gas. A summary of the observed quantities is given in \autoref{tab:observ}.

\renewcommand{\arraystretch}{1.5}
\begin{table*}
	\centering
	\caption{Observed properties of clouds from \citep{Noon2023} including the cloud identifier (Cloud ID), the median and peak \hi{} column density (\nhi{}), the median and peak CO brightness temperature (\wco{}), the observed CO line luminosity, \lco\, the average FWHM non-thermal linewidth ($\Delta v$), the radius of the clouds (R) as measured from the \hi\ observations and the atomic mass (M$_\mathrm{at}$).}
 \label{tab:observ}
 
	\begin{tabular}{ccccccccc}
		\hline\hline
        Cloud & $N_\mathrm{HI, median}$ & $N_\mathrm{HI, peak}$ & $W_\mathrm{CO, median}$ & $W_\mathrm{CO, peak}$ & \lco\ & $\Delta v$ & R & M$_\mathrm{at}$ \\
         & (cm$^{-2}$) & (cm$^{-2}$) & (K km s$^{-1}$) & (K km s$^{-1}$) & (K km s$^{-1}$ $\cdot$ pc$^2$) & (km s$^{-1}$) & (pc) & (M$_\odot$) \\
        \hline\hline
       	C1 & $5.09 \times 10^{19}$ & $1.78 \times 10^{20}$ & $0.82$ & $3.71$ & $90.92$ & $3$ & 15 & 342 \\ \hline
        C2 & $5.00 \times 10^{19}$ & $2.29 \times 10^{20}$ & $0.68$ & $3.20$ & $131.81$ & $6$ & 20 & 721 \\
        \hline
\end{tabular}
 \end{table*}

Since \htwo{} gas is difficult to directly observe in ISM environments, CO is commonly used as a tracer. Adopting a conventional CO-to-\htwo{} conversion factor of $X_\mathrm{CO} = 2 \times 10^{20}$ cm$^{-2}$ (K km s$^{-1}$)$^{-1}$ \citep[e.g.,][]{Bolatto2013a}, the inferred molecular hydrogen column densities of the Galactic wind clouds are $\sim$ 0.5-7 $\times 10^{20}$ cm$^{-2}$. The atomic hydrogen column densities derived from the 21 cm line are of comparable order, $\sim$ a few $\times 10^{20}$ cm$^{-2}$. This standard conversion, however, may not apply in the extreme environment of the GC wind, where clouds are likely exposed to elevated dissociation and ionisation rates that can modify the CO-to-\htwo{} ratio. This means the true \htwo\ content could be significantly different than implied by the standard disc conversion factor.

Although $X_\mathrm{CO}$ in the wind is uncertain, detecting any CO where there is such low \nhi\ is unusual. In the Galactic disc, CO emission is generally confined to sightlines where the atomic hydrogen columns exceed $\sim 10^{21}$ cm$^{-2}$ \citep[e.g.,][]{Rachford2009, Krumholz2009, Lee2015}. This threshold arises because \htwo{} and CO are both susceptible to photodissociation by the interstellar UV radiation field, and require substantial shielding by dust and atomic gas columns of the order $\sim 10^{21}$ cm$^{-2}$ to persist. Yet, the clouds analysed in \cite{DiTeodoro2020} and \cite{Noon2023} show detectable CO emission at \nhi{} an order of magnitude below this shielding limit. 

To explain these unusual observations, \cite{Noon2023} propose a scenario in which these clouds originate from pre-existing molecular clouds in the Galactic disc that have been swept up in the nuclear wind. Initially, these clouds are expected to consist of a dense molecular core embedded in a diffuse atomic envelope. Upon interacting with the hot and fast nuclear wind, the outer atomic layers would be rapidly stripped via a combination of hydrodynamic ablation, turbulent mixing, and thermal evaporation driven by the large temperature contrast between the cool cloud and the ambient hot medium \citep[e.g.,][]{Klein1994, Cooper2009}. This process preferentially removes the diffuse, low-density envelope, while the dense molecular core remains intact, protected by its higher column density \citep{Tonnesen2009}. The result is a molecular-dominated structure with an anomalously low \hi{} column, consistent with the observed properties. 

In this work, we test the viability of this scenario using a set of one-dimensional, equilibrium and non-equilibrium chemical models. Our goal is to determine whether the observed CO emission and \hi{} column densities can be reproduced under the assumption that the clouds originate in a molecular-dominated state and evolve under conditions representative of the nuclear wind. In particular, we seek to reproduce three key observables: the mean \hi{} column density \nhi{}, the integrated CO($2\to 1$) line luminosity \lco~(defined as the velocity- and area-integrated CO brightness temperature), and the overall cloud radius $R$, which we take to be the average of the major and minor axes of the the cloud in the \hi{} zeroth moment map. We compute all of these quantities assuming a distance of $8.2$ kpc \citep{Gravity2019}. Because our study employs a one-dimensional model, we focus on these bulk, integrated characteristics rather than attempting to reproduce the clouds' detailed substructure.

\subsection{Chemical modelling with \desp{}}
\label{ssec:despotic}

\subsubsection{The \desp{} model}

We construct chemical models for the observed clouds using \desp{} \citep{Krumholz2014}. \desp{} is a spherically-symmetric radiative transfer and chemistry code that calculates the chemical, thermal, and atomic/molecular excitation state of cool interstellar clouds, either in steady state or as a function of time. It predicts line emission using an escape probability formalism, and includes a wide range of heating and cooling processes, including cooling by line emission, heating by cosmic rays and the grain photoelectric effect, and dust-gas energy exchange at high densities. For the purposes of this work we compute chemistry using the H-He-C-O network of \citet{Gong2017}, assuming gas phase elemental abundance ratios $x_\mathrm{He} = 0.1$, $x_\mathrm{C} = 1.6\times 10^{-4}$, $x_\mathrm{O} = 3.2\times 10^{-4}$, and $x_\mathrm{Si} = 1.7\times 10^{-6}$, where we use the notation $x_\mathrm{A}$ to denote the ratio of the number density of some species A to the number density of H nuclei\footnote{To avoid confusion in what follows, we will always use the elemental symbol superscripted by 0 to denote the neutral atomic form of a species, while the elemental symbol with no superscript indicates all nuclei of that species regardless of chemical state. Thus for example $x_{\mathrm{C}^0}$ indicates the abundance of neutral atomic carbon, while $x_\mathrm{C}$ indicates the abundances of all carbon nuclei in any chemical state -- ionised, neutral, bound into CO or other molecules, etc. Similarly, we will use the notation \hi~when we are referring to the 21 cm line or quantities derived from it such as the observed column density \nhi, but we will use H$^0$ when referring to neutral atomic hydrogen otherwise.};
we further assume that the ratio of ortho- to para-H$_2$ is 0.25. We adopt the default \desp{} set of parameters for the dust, including the gas-dust collisional coupling coefficient, dust spectral index, and dust opacities per unit mass averaged over various frequency ranges -- see \citet{Krumholz2014} for a discussion of \desp{}'s parameters and their default values.

We model our clouds in \desp{} using the ``zoned cloud'' approach described in \citet{Narayanan2017}. In this approach we discretise a spherical cloud into a series of nested zones, each with its own mean density, column density to the cloud surface, and time-evolving temperature, chemical, and excitation state. Zones closer to the cloud centre are shielded against the interstellar radiation field by zones closer to the surface. We note that this model does not include collisional dissociation of \htwo{} by hot gas; however, \citet{Vijayan2024a} show that, for the H$^0$-to-\htwo{} transition, collisional effects are unimportant compared to photon-mediated ones.

\subsubsection{Cloud structure}

Since clouds in \desp{} are spherically symmetric, their physical state is fully characterised by their density profiles and linewidths. We take the latter from observations (\autoref{tab:observ}). The former is not easily observed given the constraints of two-dimensional observations, and so for simplicity we adopt a shallow powerlaw profile consistent with observations of giant molecular clouds in the disc, which have roughly constant column densities \citep[e.g.,][]{Chevance2022}, implying $n_\mathrm{H}(r)\propto r^{-1}$, where $n_\mathrm{H}$ is the number density of H nuclei. In \aref{app:constant} we present results for the case of a flat density profile for comparison.

Given our chosen density profile, the physical properties of a cloud are fully characterised by its mean volume density $\langle n_\mathrm{H}\rangle = 3 R^{-3} \int_0^R r^2 n_\mathrm{H}(r) \, dr$ and mean column density $\langle N_\mathrm{H}\rangle = 4 R^{-2} \int_0^R r^2 n_\mathrm{H}(r) \, dr$, where $R$ is the cloud radius. It follows that the outer radius of the cloud is
\begin{equation}
    R = 3 \langle N_\mathrm{H}\rangle / 4\langle n_\mathrm{H}\rangle. 
    \label{eqn:R}
\end{equation}
In a \desp{} zoned cloud, we characterise the zones by their column density to the cloud surface, where for a zone at a distance $r$ from the cloud centre we have $N_\mathrm{H} = \int_{r}^R n_\mathrm{H}(r') \, dr'$. It is again straightforward to show that the radius of a zone is related to this column density and the mean column density as
\begin{equation}
    r = R \exp\left(-2\frac{N_\mathrm{H}(r)}{\langle N_\mathrm{H}\rangle}\right),
\end{equation}
and the volume density in a zone is related to this column density as
\begin{equation}
    n_\mathrm{H}(r) = \frac{2}{3} \langle n_\mathrm{H}\rangle \exp\left(2\frac{N_\mathrm{H}(r)}{\langle N_\mathrm{H}\rangle}\right).
    \label{eqn:nh}
\end{equation}
Thus our numerical treatment of cloud structure is fully specified by the mean volume and column densities $\langle n_\mathrm{H}\rangle$ and $\langle N_\mathrm{H}\rangle$ and the choice of $N_\mathrm{H}$ values for the different zones.

To ensure that we have sufficient resolution to capture both the transition from H$^0$- to \htwo-dominated and from C$^+$- to CO-dominated composition across a zoned cloud, we employ 32 zones divided into two radial regimes. We place the innermost zone at a column density $N_\mathrm{max} = 10^{0.5} \langle N_\mathrm{H}\rangle$, so that for even the lowest mean column density cases we consider the column density is high enough that the core becomes molecular. We place the outermost zone at a fixed column density $N_\mathrm{min} = 10^{17}$ cm$^{-2}$, small enough to be transparent to far-ultraviolet (FUV) photons and thus fully atomic. We then define a transition column density $N_\mathrm{trans} = N_\mathrm{min} + (14/32)(N_\mathrm{max} - N_\mathrm{min})$, and in the inner 45\% of the zones (zones 1-14) we place the zones linearly spaced in column density from $N_\mathrm{trans}$ to $N_\mathrm{max}$, while in the outer 55\% of the zones (zones 15 - 32), we space the zones logarithmically from $N_\mathrm{min}$ to $N_\mathrm{trans}$. This strategy ensures that the core is sufficiently resolved to capture the CO-dominated zone and that the outer layers sufficiently capture the H$^0$ $\rightarrow$ \htwo{} transition. We demonstrate in \aref{app:zc} that this number of zones and zone placement strategy yields converged results for our quantities of interest.

\subsubsection{Parameterising radiation and ionisation in the nuclear wind}
\label{sss:rad}

The physical conditions within the Galactic Centre (GC) wind remain poorly constrained, particularly the strength of the interstellar radiation field and the cosmic ray ionisation rate. In \desp{} these are parameterised by $\chi$, the ultraviolet radiation field intensity normalised to the \citet{Draine1978} field, and $\zeta$, the primary ionisation rate due to cosmic rays (or X-rays). These parameters critically influence cloud thermodynamics, chemistry, and the ability of gas to self-shield against dissociating and ionising photons. 

In the absence of direct measurements, we explore five plausible environmental scenarios, designed to span the expected range of conditions. The clouds we are modelling are estimated to be $\gtrsim  1$ kpc from the GC \citep{DiTeodoro2018}, so we choose $\zeta$ and $\chi$ values that reflect diffuse disc or slightly enhanced diffuse disc conditions. Our ``R1CR1'' scenario adopts values consistent with typical Milky Way disc environments: $\chi = 1$ and $\zeta = 3 \times 10^{-17}$ s$^{-1}$ H$^{-1}$ \citep[e.g.,][]{Webber1998, VanDerTak2000, Obolentseva24a}. The ``R1CR2'' and ``R1CR3'' cases retain the fiducial radiation field but increase the ionisation rate to $\zeta = 9 \times 10^{-17}$ s$^{-1}$ H$^{-1}$ and $\zeta = 2.7 \times 10^{-16}$ s$^{-1}$ H$^{-1}$, respectively. We further consider two cases with elevated radiation fields, $\chi = 10$: the ``R2CR1'' scenario combines $\chi = 10$ with $\zeta = 3 \times 10^{-17}$ s$^{-1}$, while the ``R2CR3'' scenario has $\chi = 10$ and $\zeta = 2.7 \times 10^{-16}$ s$^{-1}$ H$^{-1}$. We summarise these cases in \autoref{tab:RFIR}.

Note that we do not consider higher ionisation rates proposed in some GC/CMZ studies \citep[e.g.,][]{Le-Petit16a}. \citet{Krumholz2023} have shown that, while ionisation rates $\sim 10^{-15}$–$10^{-14}$ s$^{-1}$ H$^{-1}$ are possible near particular local sources of cosmic rays, maintaining such a high \textit{mean} ionisation rate requires substantially more energy input than the supernova rate of the GC can plausibly supply (see also the later work by \citealt{Ravikularaman25a}, who reach much the same conclusion.) 

To avoid clutter, in the main sections of this paper we will present results primarily for the R1CR1 case. We present results for the other environments explored in \aref{app:enviro}, where we show that our major qualitative results are insensitive to the choice of radiation field. 

 \begin{table}
	\centering
	\caption{The radiation field strength ($\chi$) and ionisation rate ($\zeta$) for the five environments explored.}
 \label{tab:RFIR}
	\begin{tabular}{lcc}
		\hline\hline
         Case&$\chi$&$\zeta$\\
         &&[s$^{-1}$ H$^{-1}$]\\
        \hline\hline
		R1CR1 & $1$ & $3 \times 10^{-17}$ \\ 
        R1CR2 & $1$ & $9 \times 10^{-17}$\\
        R1CR3 &$1$ & $2.7 \times 10^{-16}$\\
        R2CR1 & $10$ & $3 \times 10^{-17}$ \\ 
        R2CR3 &$10$ & $2.7 \times 10^{-16}$\\
		\hline
\end{tabular}
 \end{table}

%%%%%%%%%%%%%%%%%%%%%%%%%%

\section{Equilibrium models}
\label{sec:equil}

Our ultimate goal is to determine under what circumstances an interstellar cloud will show the unusual combination of \hi{} column density, \nhi{}, and CO ($J=2\to 1$) integrated line luminosity, \lco\, that \cite{Noon2023} measure for nuclear wind clouds. We begin our exploration toward this goal by considering models in which we assume that the clouds are in chemical and thermal equilibrium, before moving on to non-equilibrium models in \autoref{sec:non-equ}.

\subsection{Method}
\label{sec:equ-met}

To test whether there are equilibrium configurations that can reproduce the observations, we use \desp~to compute a grid of equilibrium models for clouds with mean column densities $\log(\aNH{}/\mathrm{cm}^{-2}) = 20 - 22$ in steps of $0.013$ and $\log(\anH{}/\mathrm{cm}^{-3}) = 0.1 - 3$ in steps of $0.018$, using the average FWHM measured across the CO cloudlets for C1.\footnote{A limitation of \desp's one-dimensional framework is that clouds are characterised by a single velocity dispersion independent of radius, which raises the question of whether it is better to adopt the measured velocity dispersion for CO or \hi~for this single value. We choose the latter because the velocity dispersion has minimal effects on chemistry or 21 cm emission (since the 21 cm line is optically thin), but is important when calculating the escape probability for the optically thick CO 2$\rightarrow$1 line, which determines \lco.} Note that since $R$ depends on $\anH{}$ and $\aNH{}$ (\autoref{eqn:R}), not all points in this grid produce cloud radii that are in reasonable agreement with observations; however, for the moment we do not restrict our grid to only those models for which the cloud radius matches observations, because we can gain considerable insight by exploring the parameter space more broadly.

We run each model cloud to chemical and thermal equilibrium using \desp{}'s \texttt{setChemEq} function. This calculation determines the chemical abundances of all species in our chemical network in each zone. We then compute \nhi{} and \lco\ for the equilibrium state. We compute the former by integrating the total number of neutral hydrogen atoms over the cloud volume and dividing by the projected area,  
\begin{equation}
\label{eqn:nhi}
N_{\mathrm{H\,\textsc{i}}} = \frac{1}{\pi R^2} \int_0^R 4 \pi \nH{}(r)r^2 x_{\mathrm{H}^0}(r) dr,
\end{equation}
where \nH(r) and $x_{\mathrm{H}^0}(r)$ are the number density of hydrogen nuclei and the number of neutral hydrogen atoms per H nucleus computed by \desp{}, both as a function of radius $r$ within the cloud. We compute \lco\ from the 2$\rightarrow$1 transition using the area-weighted, velocity-integrated brightness temperature, \wco, provided by the \texttt{lineLum} function in \desp{}. This function employs the escape probability formalism to calculate line emission from each zone as a function of the zone CO abundance, temperature, column density, and linewidth. We then calculate \lco\ as $L_{\rm{CO}} = W_{\rm{CO}} \pi R^2$. 

\subsection{Results for C1}
\label{sec:equ-res}

 \begin{figure}
	\includegraphics[width=\columnwidth]{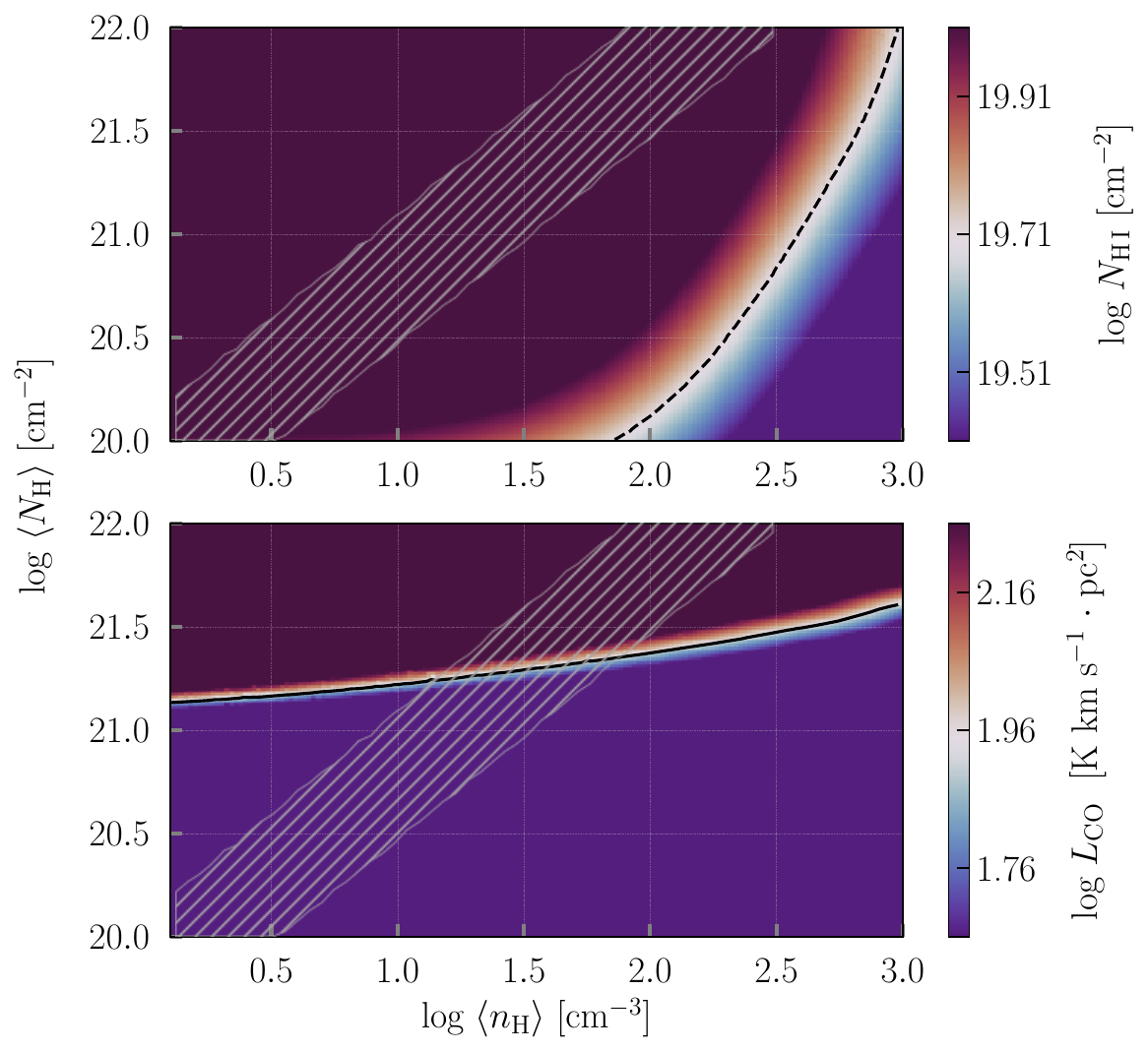}
    \caption{\hi~column density ($N_\mathrm{H~\textsc{i}}$; top panel) and integrated CO $2\to 1$ luminosity ($L_\mathrm{CO}$; bottom panel) as a function of mean hydrogen column density \aNH{} and volume density \anH{} for clouds in chemical and thermal equilibrium. In both panels, the colour scale has been set so that white corresponds to the observed values of \nhi\ and $L_\mathrm{CO}$ for C1, with the red to blue range corresponding to a factor of two variation about this value. The dashed and solid black lines in the top and bottom panels, respectively, correspond to the loci where the model-predicted and observed values of \nhi\ and $L_\mathrm{CO}$ match. The loci of (\aNH{}, \anH{}) for which the cloud radius lies within a factor of two of the observed radius of C1 is shown in the grey hatched region.}
    \label{fig:PS1}
\end{figure}

 \begin{figure}
    \includegraphics[width=7cm, height=6cm]{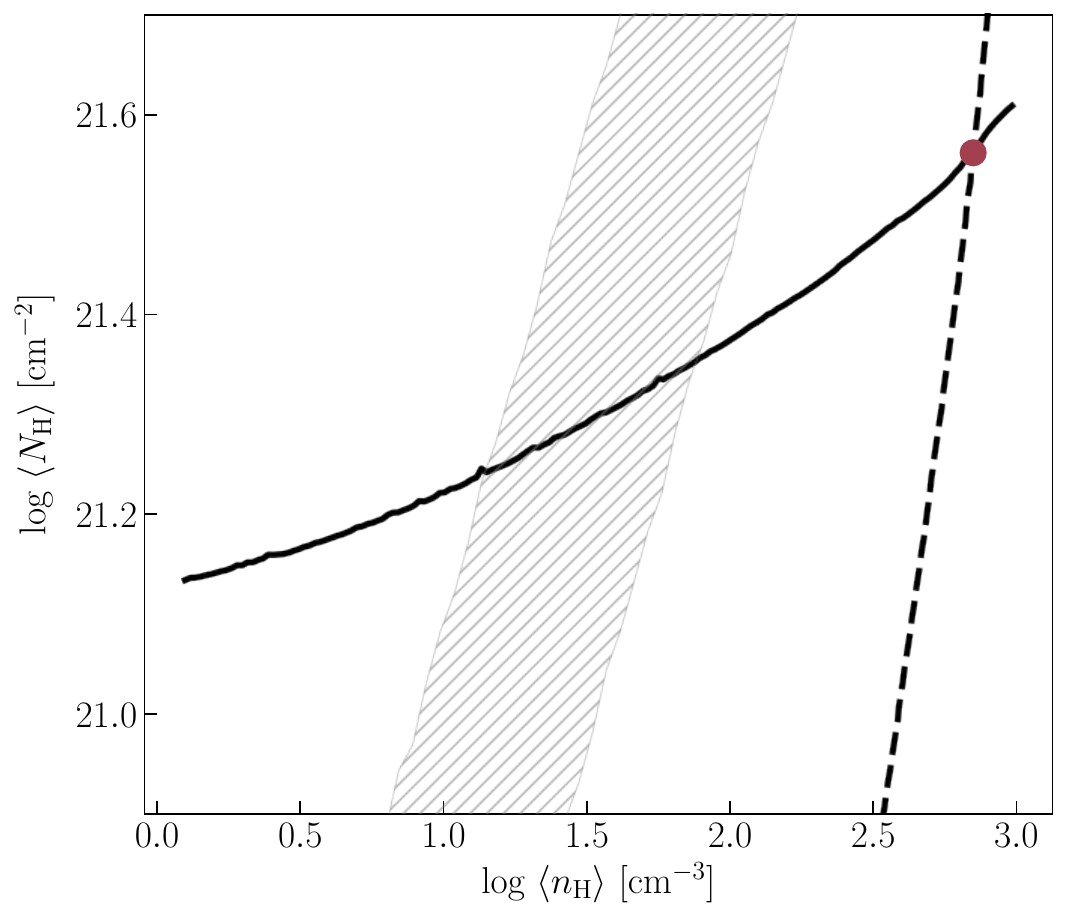}
    \caption{The same as \autoref{fig:PS1}, but now showing just the solid and dashed contour lines indicating the loci where \nhi{} (dashed) and \lco{} (solid) match the observed values for C1. The red marker indicates where the contours intersect and thus a cloud can reproduce both the observed \nhi{} and \lco{} values in equilibrium. The grey hatched region again indicates the locus where the cloud radius lies within a factor of two of the observed value.}
    \label{fig:intersect}
\end{figure}

 \begin{figure}
	\includegraphics[width=\columnwidth]{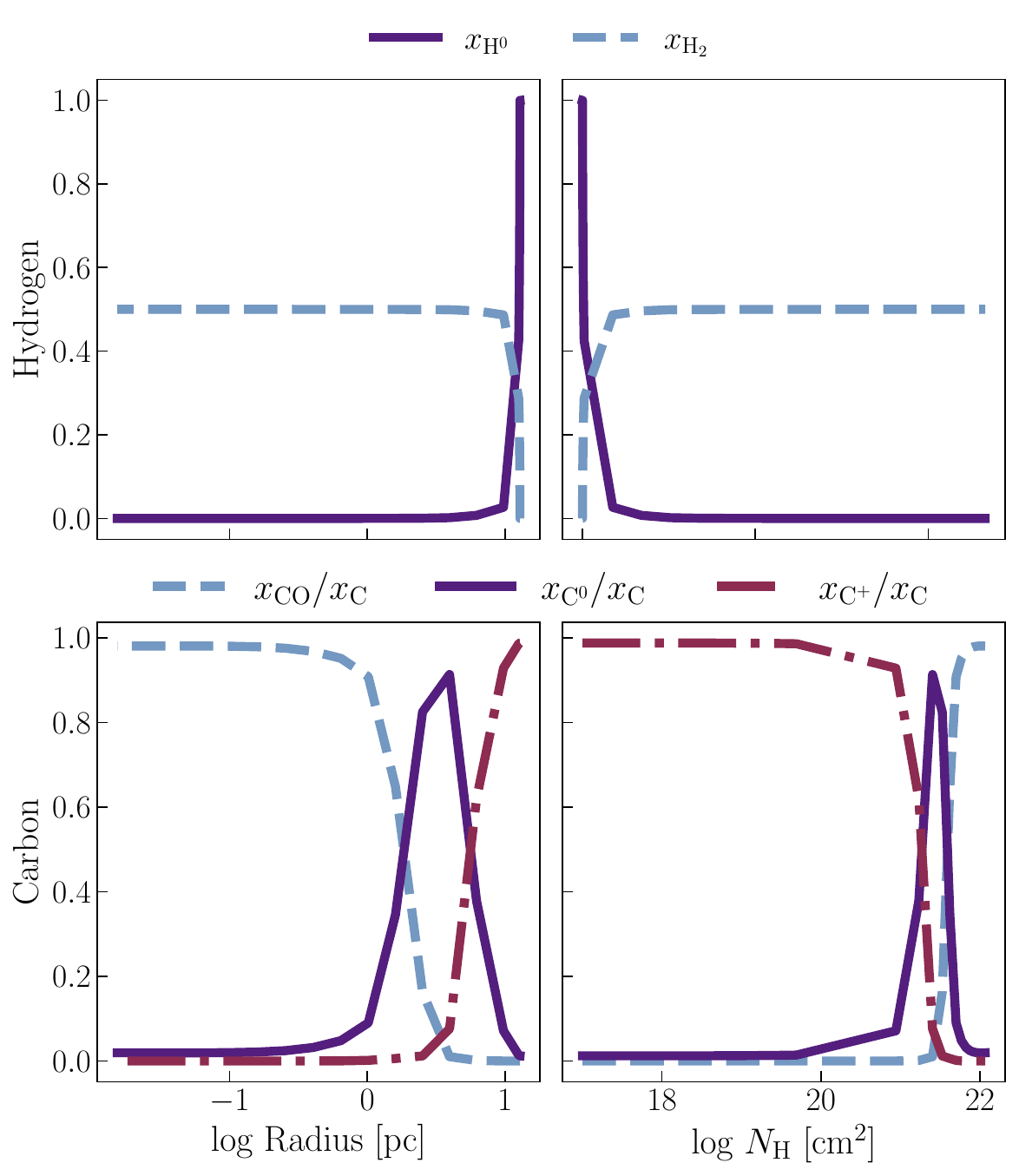}
    \caption{Abundances of select hydrogen-bearing (top row) and carbon-bearing (bottom row) chemical species as a function of cloud radius (left column) and total H column density measured from the cloud surface (right column). The case shown corresponds to $\log (\aNH{}/\mathrm{cm}^{-2}) \approx 21.57$ and $\log (\anH{}/\mathrm{cm}^{-3}) \approx 2.85$, the red point shown in \autoref{fig:intersect}. The abundances shown for the hydrogen-bearing species are normalised to the total abundance of H nuclei, while those for the carbon-bearing species are normalised to the total abundance of carbon nuclei.}
    \label{fig:abd+}
\end{figure}

We plot the values of \lco\ and \nhi{} as a function of \aNH{} and \anH{} in \autoref{fig:PS1}. In this figure, we use a narrow colour scale that shows values that match those observed for C1 in white, with the range from red to blue corresponding to a factor of two variation around these values. We also mark the (\aNH{}, \anH{}) combinations that produce a cloud radius within a factor of two of the observed radius of C1 with grey hatching. 

From \autoref{fig:PS1} we see that there is a location in \aNH{} and \anH{} space that produces a cloud that matches the observed values of \nhi{} and \lco{} in equilibrium. To highlight this, we use the data shown in \autoref{fig:PS1} to compute contours corresponding to values of \nhi{} and \lco{} that match the observed values, and we overlay these contours in \autoref{fig:intersect}. The red point in this figure highlights the point at which an equilibrium cloud can match the observations; it corresponds to $\log (\aNH{}/\mathrm{cm}^{-2}) \approx 21.57$ and $\log (\anH{}/\mathrm{cm}^{-3}) \approx 2.85$. We show the abundances of selected hydrogen- and carbon-bearing species within the cloud for this case in \autoref{fig:abd+}. The left column shows abundances as a function of radius: the cloud comprises an \htwo{}-dominated core surrounded by an H$^0$ envelope (top-left panel), while carbon exists primarily as CO in the core, transitions to neutral carbon (C$^0$) at intermediate radii, and becomes singly-ionized (C$^+$) in the outer layers (bottom-left panel). The right column shows the same abundances plotted against hydrogen column density measured from the cloud surface. This representation shows that H$^0$ dominates at low column densities (N$_{\rm{H}}$ $\lesssim$ $10^{18}$ cm$^{-2}$) while \htwo{} dominates at high columns. CO exists only in the well-shielded core (N$_{\rm{H}}$ $\gtrsim$ $10^{21}$ cm$^{-2}$) with C$^+$ dominating elsewhere.

However, we can immediately notice two problems with this equilibrium solution, one obvious and one subtle. The obvious problem is that this configuration corresponds to a cloud radius that is far from the observed cloud radius, as is visible by comparing the location of the red point in \autoref{fig:intersect} to the hatched region. The disagreement is not small: the observed cloud radius is $R \approx 15$ pc, while the identified equilibrium configuration that reproduces the observed H~\textsc{i} column and CO luminosity has $R = 1.27$ pc, more than an order of magnitude smaller, as is expected given that the mean density of the model is extremely high, close to $1000$ cm$^{-3}$.

The more subtle problem is that this equilibrium configurations requires an implausibly-large surface pressure to prevent the cloud from exploding. For our cloud density profile $n_\mathrm{H} \propto 1/r$, it is straightforward to show that the surface density is related to the mean column density and radius by $\rho_s = \aNH m_\mathrm{H}/2R$, and thus the surface pressure is
\begin{equation}
    P_s = \rho_s \sigma_\mathrm{NT}^2 = \frac{\aNH m_\mathrm{H} \sigma_\mathrm{NT}^2}{2R}.
\end{equation}
Using this relation, the surface pressure for the equilibrium configuration we have found is $P_s/k_B \approx 5 \times 10^5$ K cm$^{-3}$, which is comparable to the pressure inside a giant molecular cloud, and an order of magnitude higher than the mean pressure in the galactic disc \citep[e.g.,][]{Wolfire2003}, let alone the more diffuse halo. Assuming the cloud is confined by hot gas at a temperature of $\sim 10^6$ K, as inferred for the Milky Way halo \citep[e.g.,][]{Wang05a, Hagihara10a}, this would require the hot gas to have a number density $n_\mathrm{H}\sim \mathrm{few}\times 0.1$ cm$^{-3}$, which is $\approx 1-2$ orders of magnitude larger than the densities 1-2 kpc from the Galactic centre (the approximate location of C1) inferred from X-ray measurements \citep{Bregman2007, Gupta2012, Miller2013} or expected based on simulations of Milky Way-like galaxies \citep[e.g.,][]{Vijayan2024b, Vijayan2025}. A density this high for the hot gas would dramatically over-produce the observed X-ray brightness of the nuclear wind.

\subsection{Results for C2}
\label{sec:equil-c2}
We now repeat the procedure described in \autoref{sec:equ-met} using the linewidth measured for C2 rather than C1, and as in \autoref{sec:equ-res} we identify the region of parameter space that reproduces C2's observed combination of \nhi{}, \lco{} and radius. As with C1, we find that there is a point in $(\aNH{}, \anH{})$ space that reproduces the observed \nhi{} and \lco{} values -- $\log (\aNH{}/\mathrm{cm}^{-2}) \approx 21.41$ and $\log (\anH{}/\mathrm{cm}^{-3}) \approx 2.51$ -- but that this solution corresponds to a cloud radius substantially smaller than what is observed. Also as with C1, this equilibrium configuration requires a surface pressure far in excess of plausible confinement conditions. Thus our results for C2 are qualitatively the same as for C1: in order to reproduce the observed CO luminosity and mean \hi{} column density, a cloud would need to be an order of magnitude more compact and much denser than is observed, and would need to be confined by an implausibly high external pressure.

These problems suggest that the equilibrium configurations we have found are not plausible models for the observed clouds, and motivate the search for non-equilibrium solutions.

%%%%%%%%%%%%%%%%%%%%%%%%%%

\section{Non-equilibrium models}
\label{sec:non-equ}

\subsection{General considerations}

To investigate if the observed conditions are reproducible for a cloud that is not in chemical equilibrium, we begin by noting an important feature of the equilibrium results that is visible from \autoref{fig:PS1}. Over the range of parameters for which the cloud radius is reasonably close to the observed one, there are equilibrium configurations where the CO luminosity is also in reasonable agreement with observations (i.e., the solid line in the lower panel of \autoref{fig:PS1} passes through the hatched region), but for all such configurations the equilibrium H~\textsc{i} column density is too high. This is consistent with our expectations based on observations in the Galactic Plane, where bright CO is universally accompanied by large \hi{} columns \citep[e.g.][]{Lee2015}, and suggests that we consider non-equilibrium configurations where hydrogen atoms are temporarily locked in a different chemical state -- either H$^+$ or H$_2$ -- and have not yet had time to reach chemical equilibrium.

The former of these possibilities seems implausible on the basis that the timescale for H$^+$ to convert to H$^0$ is too short. The recombination timescale is $t_\mathrm{rec} = 1 / \alpha_\mathrm{B} n_e$, where $n_e$ is the electron density and $\alpha_\mathrm{B} \approx 3\times 10^{-13}$ cm$^{-3}$ s$^{-1}$ is the case B recombination rate coefficient. In the loci of $\anH$ and $\aNH$ where clouds can reproduce the observed CO signal, number densities are $\anH \sim 10-100$ cm$^{-3}$, and if this gas is not seen in the \hi~21 cm line because it is predominantly ionised, $n_e \sim \anH$, then the timescale over which it will become neutral is $t_\mathrm{rec} \sim 1-10$ kyr. This is orders of magnitude smaller than the $\sim 1$ Myr transit time of the cloud out of the disc.

Thus the remaining possibility is that the cloud could be in the form of H$_2$ that is in the process of dissociating into \hi. The timescale here is naturally longer because, as discussed in \citet{Noon2023}, a dissociating photon flux corresponding to $\chi = 1$ will require $\sim$Myr timescales to dissociate an H$_2$ column comparable to that of C1 or C2. This motivates us to explore a scenario where the observed \hi~column of these clouds is low because much of the hydrogen is locked in an out-of-equilibrium reservoir of H$_2$. A plausible way for this to occur would be if the cloud begins in a state of equilibrium before undergoing an event, such as being swept up by a wind, that forces it out of equilibrium by stripping away its outer layers, leaving gas that was previously shielded exposed to dissociating FUV photons.

We idealise this process by starting with a cloud in equilibrium and then instantaneously removing some of its outer, diffuse layers whilst maintaining the initial species abundances and gas temperatures across the remainder of the cloud. The stripping forces the cloud into a non-equilibrium state, from which we evolve it forward in time. We note that instantaneous stripping is almost certainly an approximation, since the timescales of mechanical acceleration and stripping are likely not too dissimilar from the chemical evolution timescale. Following this process fully self-consistently would require full time-dependent photo-chemical simulations of a cloud being entrained in a wind.\footnote{To date the published simulations that come closest to this are those of \citet{Girichidis2021}, but these begin from an atomic cloud and then consider production of molecular material by compression -- the opposite of the scenario of interest here -- and rely on a parameterised shielding prescription rather than full time-dependent radiative transfer.} Our goal here is simply to investigate a proof of concept that a cloud subject to such stripping of its outer layers is a plausible scenario to explain the unusual chemistry observed in the Galactic centre wind clouds. 

\subsection{Method}
\label{sec:noneq-meth}

\subsubsection{Implementation of stripping}
\label{sec:noneq-strip}

We define stripping as removing a portion of a cloud's outer layers. For a cloud with a density profile $n_\mathrm{H} \propto 1/r$ as we have assumed, one can immediately show that $\aNH$ is independent of the outer radius $R$, so the effect of stripping in terms of our cloud-defining parameters $\anH$ and $\aNH$ is to increase $\anH$ and decrease $R$ by the same factor, while leaving $\aNH$ unchanged.

To model stripping, we therefore characterise clouds by their initial, pre-stripping and final, post-stripping mean volume densities, \anH{}$_i$ and \anH{}$_f$; removal of the diffuse outer layers increases the mean volume density of the remainder, so $\anH{}_f > \anH{}_i$. In principle this provides us with a three-dimensional parameter space of \anH$_i$, \anH$_f$, and \aNH~to explore. Even using the fact that the observed cloud radius reduces the size of the search space somewhat, carrying out full time-dependent chemical evolution of models spanning this full space is computationally prohibitive. However, we can reduce the range of models to search by selecting \anH{}$_f$ and \aNH~not only to lie in the range of acceptable radii, but also close to the locus where in equilibrium $\lco$ is close to the observed value; graphically, this is equivalent to saying that we choose \anH$_f$ to lie both within the hatched region and close the solid line in \autoref{fig:intersect}.

Our motivation for this choice is that we expect CO emission to equilibrate faster than \hi{} emission -- equilibration of the ratio of H$_2$ to H$^0$ is slowed by self-shielding of H$_2$ molecules, whereas self-shielding is unimportant for dissociation of CO. Thus we expect \lco~to be close to its equilibrium value even a short time after stripping, indicating that if we want to find non-equilibrium clouds that reproduce observations, we should look for configurations where \hi~is far from equilibrium, but CO is not. In practice we find that, because CO does not instantaneously reach equilibrium, we get better matches to observations by choosing values of \anH$_f$ that are slightly (+0.15 dex in log-space) to the right of the equilibrium \lco\ line in \autoref{fig:intersect}, so that the equilibrium CO luminosity is slightly lower than the observed value. This offset accounts for the different timescales on which \hi{} and CO re-equilibrate. Experiments with alternative values for the amount of offset from the equilibrium \lco\ line shows that this choice affects details of the range of models that we later deem valid, but does not change the qualitative results regarding the range of physical conditions that can reproduce the observations. 

\begin{figure*}
    \centering
    \begin{subfigure}[t]{0.5\textwidth}
        \centering
        \includegraphics[width=8.cm, height=8.cm]{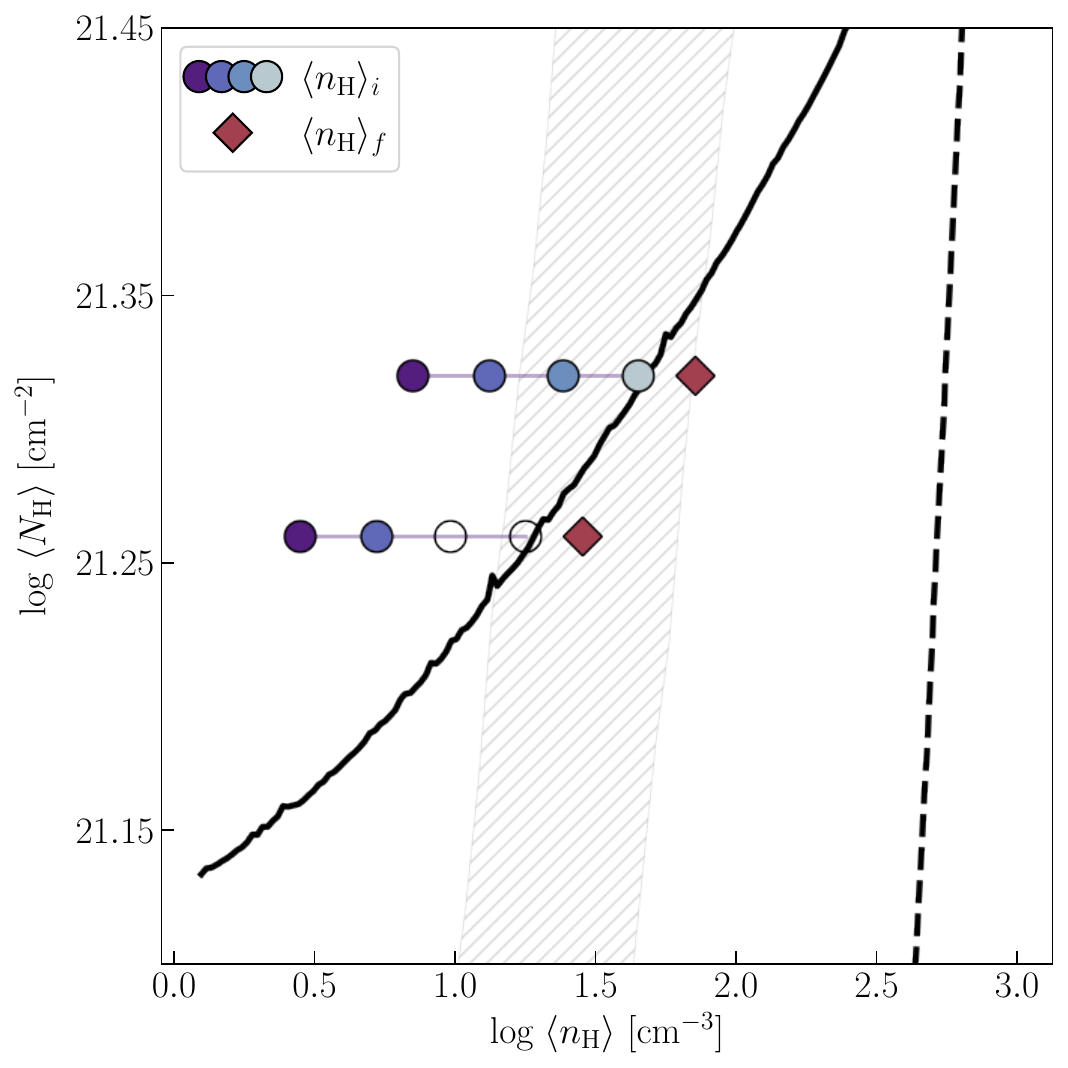}
        \caption{C1}
    \end{subfigure}%
    ~ 
    \begin{subfigure}[t]{0.5\textwidth}
        \centering
        \includegraphics[width=8.cm, height=8.cm]{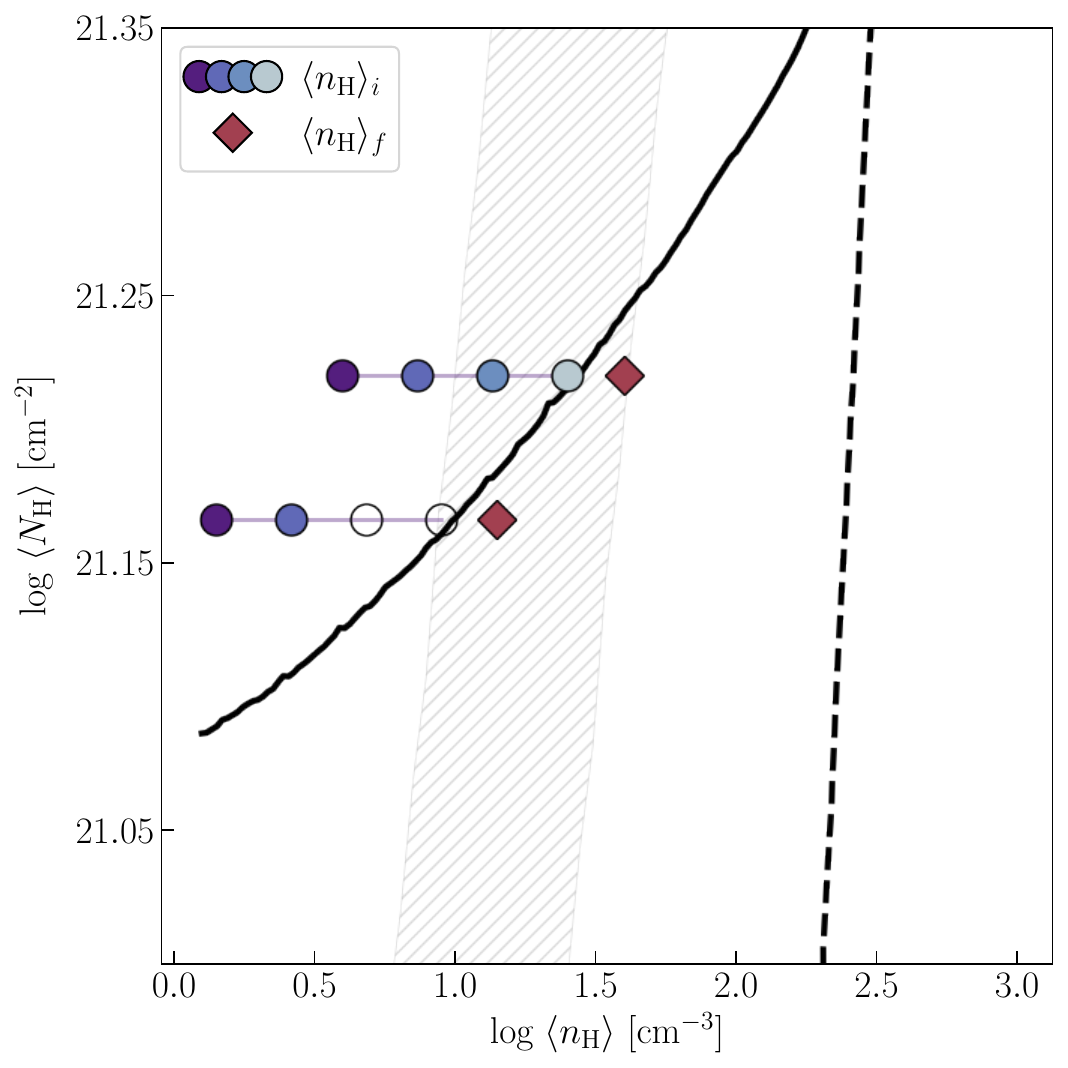}
        \caption{C2}
    \end{subfigure}
    \caption{Same as in \autoref{fig:intersect}, but with markers showing a selection of \anH{}$_i$ (circle markers) and \anH{}$_f$ (diamond markers) values. The \anH{}$_i$ values connected by the horizontal line correspond to the \anH{}$_f$ point at the same \aNH{}. The exact coordinates of the points shown are given in \autoref{tab:models}. The filled markers represent the models that successfully reproduce the observed \lco, \nhi, and radii of (a) C1 and (b) C2, while empty markers show models that do not reproduce the observations. The grey hatched regions show loci where the cloud radii are within a factor of two of the observed radii, and the solid and dashed lines show the loci for which the equilibrium values of \lco~and \nhi~match that of the observed clouds; these lines in (a) are identical to those shown for C1 in \autoref{fig:intersect}.}
    \label{fig:intersection_markers}
\end{figure*}

Thus we consider non-equilibrium clouds in which the final, post-stripped state $(\anH_f, \aNH)$ lies 0.15 dex to the right of the solid line in \autoref{fig:intersect} and within the hatched region indicating clouds whose radii are close to what we observe. The initial pre-stripped state lies somewhere to the left of this, at the same \aNH~but lower \anH. We therefore consider two $(\anH_f, \aNH)$ pairs that lie within the range of acceptable radii and 0.15 dex to the right of the solid line in \autoref{fig:intersect}, and pair these with four \anH{}$_i$ values that are uniformly spaced from 0.2 to 1 dex lower than \anH$_f$, giving us a total of eight models. We list the full set of models we explore for C1 and C2 in \autoref{tab:models}, and illustrate the parameter-space positions for each cloud in \autoref{fig:intersection_markers}.

 \begin{figure}
    \includegraphics[width=\columnwidth]{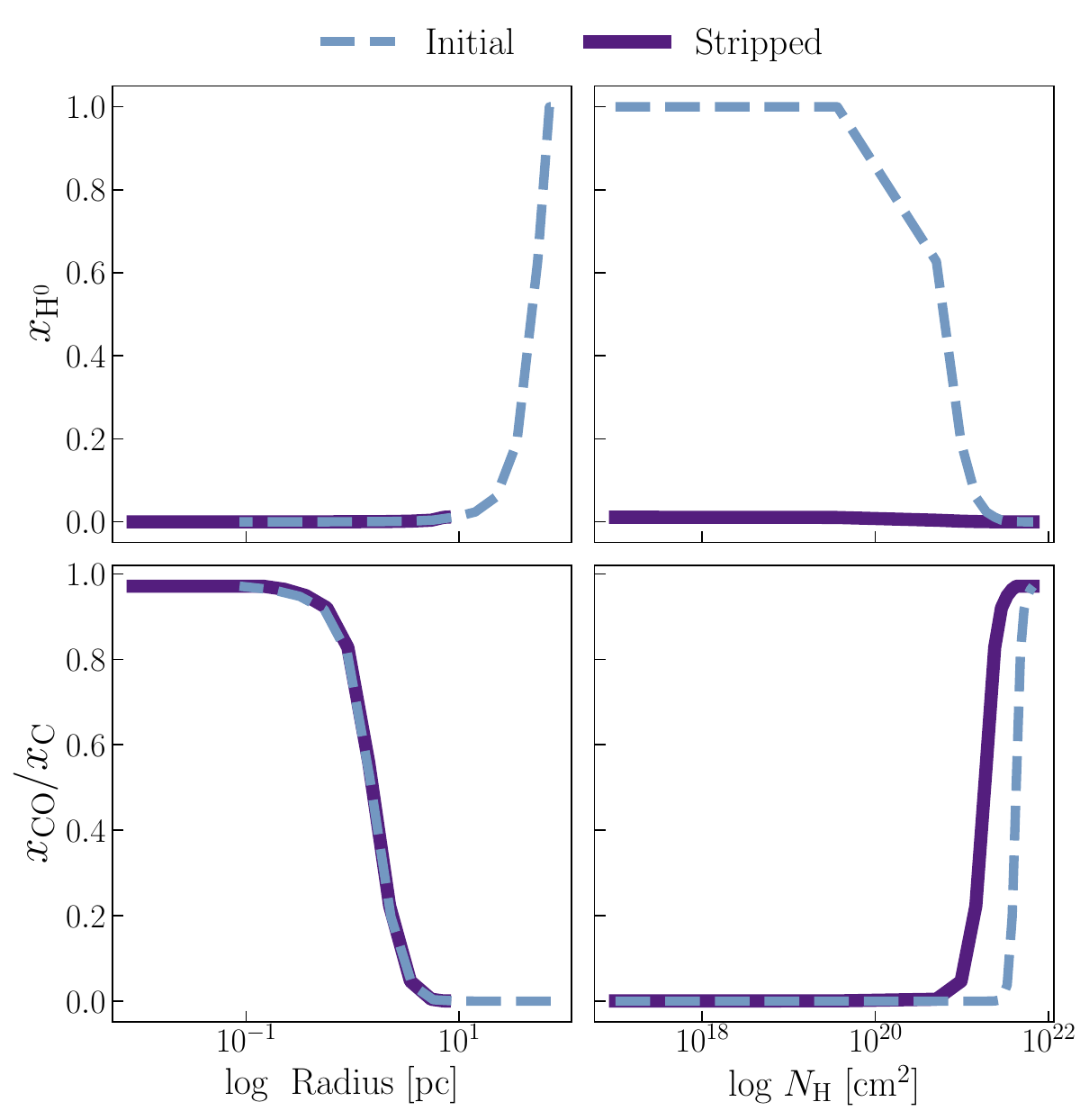}
    \caption{\hi{} (top panel) and CO (bottom panel) normalised abundances as a function of cloud radius (left column) and total H column density to the cloud surface (right column) for the C1 model $\log(\anH_i/\mathrm{cm}^{-3}) = 0.85$, $\log(\anH_f/\mathrm{cm}^{-3}) = 1.85$ and $\log(\aNH/\mathrm{cm}^{-2}) = 21.32$ (the most stripped model for row two in \autoref{tab:models}); the abundances $x_\mathrm{H^0}$ and $x_\mathrm{CO}$ are the number of H$^0$ and CO per H nucleon, while $x_\mathrm{C}$ is the number of C nuclei per H nucleon. The dashed blue line shows the initial, equilibrium abundances of the cloud before stripping, and the solid purple line shows the non-equilibrium abundances immediately after the cloud has been stripped.}
    \label{fig:abd_stripped}
\end{figure}

\begin{table}
    \centering
    \begin{tabular}{cccc}
    \hline\hline
    Cloud & $\log\aNH{}$ & $\log\langle n_\mathrm{H}\rangle_f$ & $\log\langle n_\mathrm{H}\rangle_i$ \\
    
    & [cm$^{-2}$] & [cm$^{-3}$] & [cm$^{-3}$] \\
    \hline\hline
    \multirow{2}{*}{C1} & 21.26 & 1.45 & 0.45, 0.72, 0.98, 1.25 \\
    & 21.32 & 1.85 & 0.85, 1.12, 1.38, 1.65 \\ 
    \hline
    \multirow{2}{*}{C2} & 21.17 & 1.15 & 0.15, 0.42, 0.68, 0.95 \\
    & 21.22 & 1.60 & 0.60, 0.87, 1.13, 1.40 \\

    \hline
    \end{tabular}
    \caption{Parameters of non-equilibrium chemical models explored for the R1CR1 case. Each model is characterised by a mean cloud column density \aNH{}, a mean volume density after stripping $\langle n_\mathrm{H}\rangle_f$, and a mean volume density before stripping $\langle n_\mathrm{H}\rangle_i$.}
    \label{tab:models}
\end{table}

For each model we initialise our calculations by using \desp{} to find the chemical equilibrium state for a cloud using the chosen \anH{}$_{i}$ and \aNH{} values following the same method outlined in \autoref{sec:equ-met}. We then construct interpolating functions giving the temperature and abundances of all chemical species as a function of cloud radius. To initialise the out-of-equilibrium configuration at \anH$_{f}$ and \aNH~that we will evolve in time, we use the same number of zones and zone column densities as for the initial cloud (since \aNH~is unchanged by stripping), set the density of all zones by using \anH$_{f}$ in \autoref{eqn:nh}, and set the gas temperature and chemical abundances in all zones using the interpolating functions we constructed from the pre-stripped configuration. In this way we ensure that the remaining parts of the stripped cloud begin from a density, thermal, and chemical structure identical to that of the pre-stripped cloud. We illustrate this process in \autoref{fig:abd_stripped}, which shows both the pre-stripping and post-stripping configurations for an example case corresponding to the upper left purple circular point in panel (a) of \autoref{fig:intersection_markers} ($\log(\anH_i/\mathrm{cm}^{-3}) = 0.85$, $\log(\anH_f/\mathrm{cm}^{-3}) = 1.85$ and $\log(\aNH/\mathrm{cm}^{-2}) = 21.32$). We see that stripping leaves intact the part of the cloud in which there is a transition from CO to C to C$^+$, but removes the part where the H$^0$ to H$_2$ transition occurs, so there is no H$^0$-dominated zone and H$_2$ is directly exposed to the interstellar radiation field (ISRF). For models that undergo less stripping, a thin outer layer of H$^0$ remains with abundance fractions reaching $\sim 15-20\%$ in the outer layers, thought still significantly less than in an equilibrium cloud. 

\subsubsection{Time evolution}
\label{sec:noneq-evo}
We evolve each of the models listed in \autoref{tab:models} over time using \desp{}'s $\textsc{chemEvol}$ function, which solves the system of ordinary differential equations describing the time evolution of the species abundances. We use the \textsc{gasEq} option for this calculation, which keeps the dust temperature fixed and sets the gas temperature at all times to its instantaneous equilibrium value. Neglect of dust temperature evolution is reasonable because for the conditions in which we are interested, dust-gas collisional coupling is unimportant and thus the dust temperature has no significant effect. Similarly, the assumption of instantaneous thermal equilibrium for the gas is reasonable because the thermal equilibration timescale for the gas is always short compared to the chemical evolution timescale \citep[e.g.,][]{Krumholz12e}. We evolve the clouds for one Myr, as all the models explored approach a new equilibrium state within this time-frame. 

\subsection{Results}
\label{sec:noneq-res}

 \begin{figure}
    \includegraphics[width=\columnwidth]{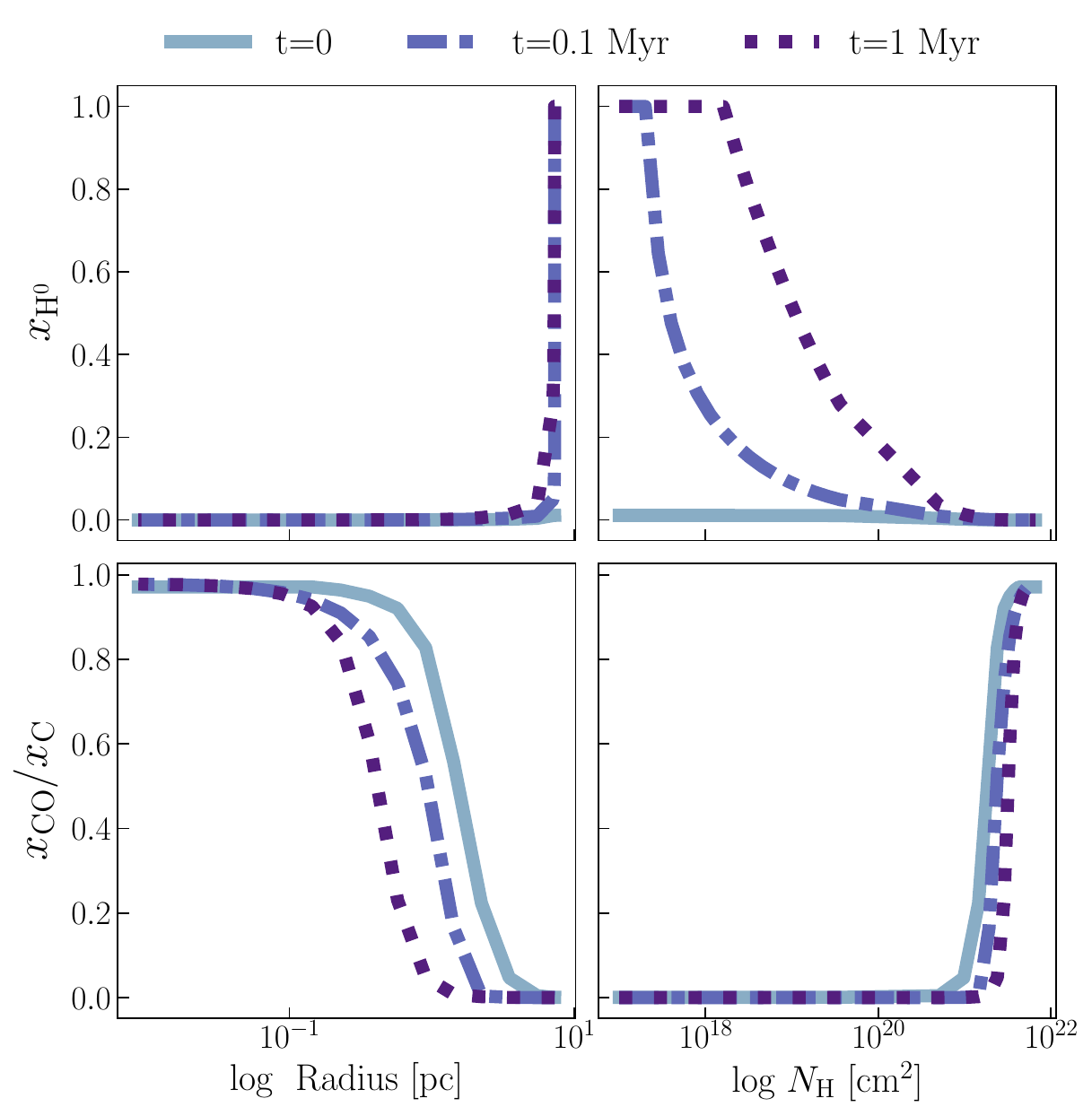}
    \caption{Same as \autoref{fig:abd_stripped}, but now showing the state produced for 0.1 Myr (dash-dotted line) and 1 Myr (dotted line) of chemical evolution after stripping. The solid blue line labelled $t=0$ is the same as the ``Stripped'' state shown in \autoref{fig:abd_stripped}.}
    \label{fig:abd_stripped_evolved}
\end{figure}

 \begin{figure}
 \includegraphics[width=\columnwidth]{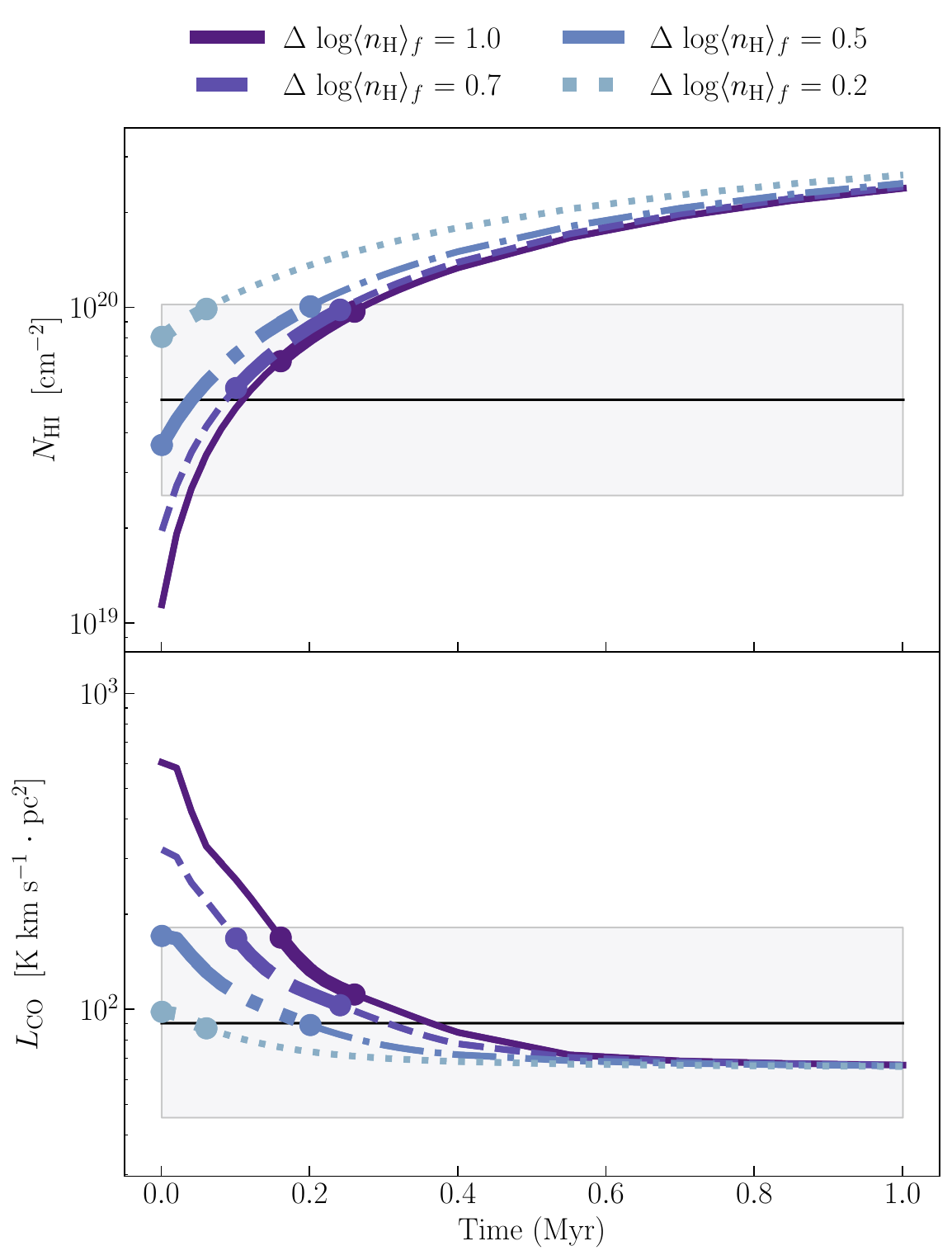}
    \caption{Time evolution of \nhi{} (top) and \lco{} (bottom) for clouds with $\log(\aNH/\mathrm{cm}^{-2}) = 21.32$ and $\log(\anH{}_f/\mathrm{cm}^{-3})=1.85$ for a range of values of $\log \anH_i$ as indicated in the legend; the quantities denoted in the legend are $\Delta \log\anH = \log(\anH{}_f/\mathrm{cm}^{-3}) - \log(\anH{}_i/\mathrm{cm}^{-3})$. The solid black line and grey shaded bands indicate the median values of \nhi{} and \lco{} observed for C1, with factors of two scatter above and below these values. The thick line segments between the two markers indicate the time intervals that the models are valid, i.e., \nhi\, \lco\, and cloud radius all differ from the observed values by less than a factor of two.}
    \label{fig:evolve_1}
\end{figure}

\autoref{fig:abd_stripped_evolved} shows an example of the results of this calculation, using the same case as shown in \autoref{fig:abd_stripped}: $\log(\anH_i/\mathrm{cm}^{-3}) = 0.85$, $\log(\anH_f/\mathrm{cm}^{-3}) = 1.85$ and $\log(\aNH/\mathrm{cm}^{-2}) = 21.32$, and with a linewidth matching that of C1. The line labelled $t=0$ corresponds to the state immediately after stripping, and the other lines show the chemical state after 0.1 and 1 Myr of evolution. We see that as the cloud evolves, the CO-dominated zone moves inward slightly due to the reduction in dust shielding, but this is a relatively minor effect: the CO-dominated zone moves inward from a few pc to $\sim 1$ pc in radius, and from $\approx 10^{21}$ cm$^{-2}$ to $\approx 2\times 10^{21}$ cm$^{-2}$ in column density (as measured from the cloud surface). By contrast, the change in hydrogen chemistry is much more dramatic. The H$_2$ at the cloud surface begins to dissociate, causing an H$^0$-dominated zone to re-appear, and this zone moves inward to $\approx 10^{18}$ cm$^{-2}$ after 0.1 Myr, and reaches to $\approx 10^{19}$ cm$^{-2}$ after 1 Myr. Thus we can immediately see that there are times during the evolution when the H$^0$-dominated zone has not yet reformed but the CO is close to its final equilibrium state, and so the cloud can be expected to have low \nhi~combined with high \lco, precisely the situation of interest to us.

We confirm this expectation in \autoref{fig:evolve_1}, where we plot the observable mean \hi{} column density \nhi{} and CO 2$\to$1 luminosity \lco~as a function of time for the four different values of \anH{}$_i$ for the same example cloud. Consistent with our expectations, we see that the CO luminosity changes by a smaller amount than \nhi{}, and that its evolution is more rapid. Even for the case with the largest stripping, the change in CO luminosity is less than a factor of 10, and this change is essentially complete after $\approx 0.5$ Myr of evolution. By contrast, the change in \nhi~after 1 Myr is more than a factor of 30, and the value is clearly still increasing. Most importantly, we see that there are indeed times when the cloud displays values of these quantities that are reasonably close to the observed ones for C1, as indicated by the grey bands in the figure.

To more broadly characterise under what circumstances we can reproduce the observed combination of \nhi~and \lco, and to measure the physical properties of the clouds during the times in their evolution when this is the case, we examine all of our simulations and flag any times when the cloud is ``valid''. We define valid to mean that the model's values of \nhi{} and \lco{} are both within a factor of two of the median observed values -- using the observed values for either C1 or C2, depending on which model we are testing -- at the same time. For the example shown in \autoref{fig:evolve_1}, we mark the interval of valid times for each model as thick lines between the markers. We find that valid times exist for all of the parameter combinations in \autoref{tab:models} except $\log(\aNH/\mathrm{cm}^{-2}) = 21.26$ with $\log(\anH_i/\mathrm{cm}^{-2}) = 0.98$ or $1.25$ for C1, and for all models except $\log(\aNH/\mathrm{cm}^{-2}) = 21.17$ with $\log(\anH_i/\mathrm{cm}^{-2}) = 0.68$ or $0.95$ for C2. These models fail because they are not stripped enough to reach \nhi~values low enough to match the observations. Nonetheless, the fact that six of our eight models for both C1 and C2 experience some point during their evolution during which their emission is a reasonably close match to that of the observed clouds is strong evidence that unusual combinations of \nhi~and \lco~such as those observed can be produced by non-equilibrium chemistry.

Moreover, these valid models resolve the problems we identified with the equilibrium solution in \autoref{sec:equ-res}. First, by construction the cloud radius is reasonably close to observations. Second, the associated pressures and densities of the non-equilibrium cloud models are much more realistic. For all valid models and times, we obtain a pressure range of $P_s / k_B = (2 - 5) \times 10^4$ K cm$^{-3}$, comparable to typical Galactic disc pressure. The corresponding number densities for a $T = 10^{6}$ K wind are $\sim 10^{-2}$ cm$^{-3}$, consistent with estimates from both X-ray observations and numerical simulations \citep{Bregman2007, Gupta2012, Miller2013, Vijayan2025}. Consequently, non-equilibrium cloud configurations for cold MW wind clouds appear more likely than equilibrium solutions. 

%%%%%%%%%%%%%%%%%%%%%%%%%%

\section{Discussion}
\label{sec:disc}
The key result we obtain from this study is that at least some of the clouds entrained in the MW's nuclear wind cannot be in chemical equilibrium. Their observed combination of \nhi{}, \lco{}, and physical size requires that they have been observed in a transient, non-equilibrium state. By exploring the space of possible non-equilibrium configurations, we identify the physical conditions most likely to produce chemical states consistent with the observations. Here we discuss the implications of these results.

\subsection{Cloud origins and timescales}

The first implication concerns the origin and evolutionary timescale of the clouds. Our results show that a cloud that begins as a typical disc molecular cloud and experiences partial removal of its low-density envelope as it becomes entrained in a wind evolves into a configuration resembling the observed MW wind clouds. In particular, the longer time taken by hydrogen relative to carbon chemistry to adjust to the new physical state after this rapid change explains the unusually low \nhi{} and detectable \lco{} that characterise the observed clouds. This in turn suggests that it must be possible for clouds that are entrained in the wind to survive, contrary to earlier theoretical expectations that they should be destroyed on short timescales \citep{Klein1994, Scannapieco2015}, and consistent with more recent studies suggesting that a combination of magnetic fields, thermal conduction, and radiative cooling can enable long-term survival of cool clouds \citep[e.g.,][]{McCourt2015,Armillotta2017, Gronke2022, Tan2024}. Confirming this result will require three-dimensional simulations including time-dependent chemistry and radiative transfer, and with enough resolution to capture the chemical transition layers between different gas phases.

However, a complete picture of the suite of principal mechanisms responsible for cold cloud survival is still to be determined. Our work suggests that chemical constraints derived from realistic, time-dependent photochemistry could provide important crucial constraints on this problem.

\subsection{CO-H$_{\rm{2}}$ conversion factor}

A second important application of our results is that they allow us to estimate the CO-to-\htwo{} conversion factor, $X_{\rm CO}$, for the wind clouds. This conversion factor is crucial for estimating the molecular mass outflow rate of the MW, and for other galaxies where winds are observed in CO transitions (e.g., M82 -- \citealt{Leroy2015, Yuan2023}). In the context of our models, we define 
\begin{equation}
    X_\mathrm{CO} = \frac{M_\mathrm{H_2}}{m_\mathrm{H_2} W_\mathrm{CO}} = \frac{\int_0^R 4\pi n_\mathrm{H} x_\mathrm{H_2}(r) r^2 \, dr}{m_\mathrm{H_2} L_\mathrm{CO}}.
\end{equation}
Here the numerator is the total mass of molecular hydrogen within a cloud, $m_\mathrm{H_2}$ is the mass of a single H$_2$ molecule, $R$ is the cloud outer radius, and $L_\mathrm{CO}$ is the CO luminosity, which we can compute for any CO transition of our choice. As is usual, we express the luminosity as the product of the cloud area $A=\pi R^2$ and the velocity-integrated brightness temperature, $L_\mathrm{CO} = W_\mathrm{CO} A$, with units of K km s$^{-1}$ pc$^{2}$, so that $X_\mathrm{CO}$ has the customary units of cm$^{-2}$ / (K km s$^{-1}$)$^{-1}$. We compute this quantity for any valid non-equilibrium model identified in \autoref{sec:noneq-res}.

We collect all of the ``valid'' time snapshots for the non-equilibrium R1CR1 models and compute $X_\mathrm{CO}$ in both the CO $2\to 1$ and $1\to 0$ lines for each of them. We report the full range of $X_\mathrm{CO}$ values for all these snapshots in \autoref{tab:XCO}. We find $X_\mathrm{CO}$ values $\approx (0.7-5)\times 10^{21}$ cm$^{-2}$ (K km s$^{-1}$)$^{-1}$ in both lines, significantly larger than the values of $\approx (2-4)\times 10^{20}$ cm$^{-2}$ (K km s$^{-1}$)$^{-1}$ typically assumed in Galactic discs, and near the upper end of the range assumed in previous analysis of these clouds derived from equilibrium models \citep{DiTeodoro2020, Heyer2025}.

 \begin{table}
    \centering
    \begin{tabular}{ccc}
    \hline\hline
    Cloud &$X_\mathrm{CO(2\to 1)}$ & $X_\mathrm{CO(1\to 0)}$ \\
    & (cm$^{-2}$) (K km s$^{-1}$)$^{-1}$ & (cm$^{-2}$) (K km s$^{-1}$)$^{-1}$ \\
    \hline\hline
    
    C1 & $1.01-4.33\times10^{21}$ & $0.69-2.78\times10^{21}$\\ 
    \hline
    
    C2 & $1.20-6.16\times10^{21}$ & $0.86-3.96\times10^{21}$\\
    \hline
    \end{tabular}
    \caption{The range of $X_\mathrm{CO(2\to 1)}$ and $X_\mathrm{CO(1\to 0)}$ values for valid non-equilibrium models for C1 and C2 in the R1CR1 environment.}
    \label{tab:XCO}
\end{table}

Recently, \cite{DiTeodoro2026} analysed a sample of 16 wind CO clouds using \desp\ one-zone (rather than multi-zone) equilibrium models. They do find equilibrium models that can reproduce the observations reasonably well, and further find for these models similarly elevated $X_\mathrm{CO}$ values, $X_\mathrm{CO(2\to 1)} = (0.6-2)\times 10^{21}$ cm$^{-2}$ (K km s$^{-1}$)$^{-1}$. Given that we find that equilibrium models cannot reproduce the observed CO and \hi~data, but nonetheless we obtain similar $X_\mathrm{CO}$ values from our non-equilibrium models, it is important to understand how our results differ from theirs.

Their approach was similar to ours in that they attempted to reproduce the measured values of \wco, \nhi, and cloud size, but with one key difference: the cloud radii that \citet{DiTeodoro2026} attempt to reproduce are those measured from the CO emission, which yields substantially smaller sizes ($\approx 1-2$ pc) than those inferred from the atomic gas ($\approx 15-20$ pc), which we use here. Arguments can be made for either choice. Using the smaller size of individual CO clumps for the purpose of modelling in \desp\ reflects the fact that the observed clouds have multiple CO clumps within the one \hi\ envelope. Moreover, \citeauthor{DiTeodoro2026} lack spatially resolved \hi\ data for most of their sample, making a CO-based approach necessary. By contrast, our choice to use the \hi-based size is motivated by the fact that the atomic envelope around the molecular gas plays a crucial role in shielding the molecular core, and it is therefore critical to choose a size that captures the extent of this envelope. And because we are modelling only two of the sixteen clouds analysed by \citeauthor{DiTeodoro2026}, we do have access to spatially resolved \hi\ data.

Regardless of the arguments for one choice versus the other, \autoref{sec:equ-res} makes it clear that this choice has significant consequences: if we were to reduce the target radius for C1 from $\approx 15$ pc (observed in the \hi{} data) to $\approx 1-2$ pc (the observed size of the CO-bright region), the equilibrium solution shown in \autoref{fig:intersect} would become acceptable. Thus it is the fact that we determine cloud radii from the resolved \hi\ emission that rules out the possibility of valid equilibrium models for us. Ultimately, however, we caution that both the \citeauthor{DiTeodoro2026} equilibrium models and our non-equilibrium models represent approximations of a complex reality: observations reveal clumpy molecular structures with multiple CO clumps embedded within \hi\ envelopes, a configuration that it is not possible to fully capture in \desp{}'s one-dimensional framework. Further exploration will require three-dimensional simulations combining hydrodynamics, time-dependent chemistry, and radiative transfer of FUV radiation.

Given this analysis, it might at first seem surprising that we and \citet{DiTeodoro2026} obtain similar values of $X_\mathrm{CO}$. However, we can understand this as arising from a common feature of the data that must be reproduced in any scenario: the unusually low \nhi~value for a cloud with a CO-bright core. In our non-equilibrium models, we explain this feature as arising because CO re-equilibrates on timescales short compared to the hydrogen chemistry following stripping, leaving a large non-equilibrium H$_2$ reservoir that has not yet had time to dissociate. By contrast, equilibrium models are able to achieve low \nhi~because their very large mean densities ($\sim 10^3$ cm$^{-3}$) allow the H$^0$ to convert to H$_2$ even where it is only weakly shielded from dissociating UV light, yielding a transition to H$_2$ at much lower \hi~column than would be possible for the lower gas densities typically found in Galactic plane clouds. Thus while these models disagree about the physical reason for there being an unusually large H$_2$ reservoir relative to the amount of \hi{} -- non-equilibrium chemistry versus exceptionally high density -- they both agree that such a reservoir of H$_2$ must be present, and this in turn implies large $X_\mathrm{CO}$. This explains why both the equilibrium models of \citet{DiTeodoro2026} and the non-equilibrium models presented here yield similar $X_{\rm{CO}}$ values, despite their differences.

\subsection{Molecular mass of the wind clouds and chemical state of the Milky Way's nuclear wind}

Using the $X_\mathrm{CO}$ values inferred from our valid models for both the $2\to 1$ and $1\to 0$ lines, we estimate the molecular masses of our two clouds as $M_{\rm{mol}} = 1.36- 8.53 \times 10^{3}$ M$_\odot$ for C1 and $M_{\rm{mol}} = 2.46 - 17.59 \times 10^{3}$ M$_\odot$ for C2; note that these figures assume that the mean mass per H$_2$ molecule is $3.9\times 10^{-24}$ g to account for the mass of He. These values are at least an order of magnitude larger than previous estimates that assumed a standard Galactic disc conversion factor, $X_\mathrm{CO} = 2\times 10^{20}$ (cm$^{-2}$) (K km s$^{-1}$)$^{-1}$, which yields $\sim 400$ M$_\odot$ for C1 and $\sim 400-600$ M$_\odot$ for C2 \citep{DiTeodoro2020, Noon2023}. By contrast \citet{DiTeodoro2026} find molecular mass ranges $M_{\rm{mol}} \approx 2.22-5.39 \times 10^{3}$ M$_\odot$ for C1, overlapping with our results, and $M_{\rm{mol}} \approx 2.98-6.48 \times 10^{3}$ M$_\odot$ for C2, which still overlaps with our results but only at the lower end of the mass range inferred here, consistent with our higher $X_\mathrm{CO}$ values. The difference in results between ours and those presented in \citet{DiTeodoro2026} for C2 likely reflects the differences in how the results are calculated and reported. In our method, we use the integrated \lco\ to define the cloud's CO emission, whereas \citet{DiTeodoro2026} use \wco. Whilst \wco\ is of the same order of magnitude between C1 and C2, this is not the case for \lco\ (\autoref{tab:observ}) -- C2 has larger \lco\ than C1 and thus permits lower-density configurations with higher $X_\mathrm{CO}$. Further, \citet{DiTeodoro2026} report median values across all clouds while we report the full range from all valid models. A combination of these two differences could result in the disagreement in results we observe for C2. Conversely, C1's lower \lco\ reduces the range of valid models, resulting in a better agreement between the two approaches. 

However, regardless of method, both our non-equilibrium models and the equilibrium models of \citet{DiTeodoro2026} point to the same conclusion: the molecular mass fraction of the Milky Way's nuclear wind is substantially larger than previous estimates suggest. While \citeauthor{DiTeodoro2026} find that the molecular fraction drops at larger distances, $\sim 1$ kpc from the plane, our results show that close to the disc where the outflow is launched, the neutral mass flux is dominated by the molecular phase, with the neutral atomic phase representing only a $\approx 10\%$ correction.

The increase in cloud mass estimates also carries with it a concomitant increase in estimate of the Milky Way wind's mass loading factor (defined as the ratio of wind mass flux to star formation rates). Previous estimates based on atomic gas suggested a wind mass flux $\approx 0.1$ M$_\odot$ yr \citep[e.g.,][]{DiTeodoro2018}, compared to a total Galactic star formation rate $\approx$ 2 $M_\odot$ yr$^{-1}$ \citep[e.g.,][]{Chomiuk2011}, and a star formation rate (SFR) $\approx 0.1$ M$_\odot$ yr$^{-1}$ in the Milky Way's CMZ \citep[e.g.,][]{Immer2012}. Thus these earlier estimates suggested a mass loading factor $\eta \sim 0.1$ or $\sim 1$, depending on whether we take the denominator to be the SFR of the whole galaxy or just the CMZ. In either case, our revised $X_\mathrm{CO}$ values suggest a factor of $\approx 10$ increase, to $\eta \sim 1$ or $\sim 10$ within $\pm 1$ kpc of the Galactic plane. Molecular outflows are commonly detected in starburst galaxies \citep[e.g.][]{Leroy2015, Walter2017, Fisher2025} with similar mass loading factors \citep{Bolatto2013b, Cicone2014}, and so our revised molecular masses for MW wind clouds bring the Galaxy's nuclear outflow more in line with this broader population. This in turn suggests that molecular gas entrainment and survival is likely a common feature of galactic winds.

%%%%%%%%%%%%%%%%%%%%%%%%%%

\section{Conclusions}
\label{sec:conc}
We have used the time-dependant photochemistry code \desp{} to model two molecular clouds in the Milky Way nuclear wind that have unusually small \hi~column densities for their high CO luminosities, suggesting a chemical state very different from that found for molecular clouds in the Galactic Plane. We have explored both chemical equilibrium and non-equilibrium scenarios to reproduce these unusual states. Our principal findings are:
\begin{enumerate}

\item The wind clouds are out of chemical equilibrium. No equilibrium model can simultaneously reproduce the observed CO luminosity, \hi{} column density, and cloud size for either cloud. Conversely, non-equilibrium models can reproduce all three observables. The most natural explanation for the observations is therefore that the observed clouds are in a transient chemical state.

\item The non-equilibrium chemistry points to a specific formation scenario whereby the clouds began as typical molecular clouds in the Galactic disc and experienced rapid, partial removal of their low-density envelopes upon entrainment in the wind. Carbon chemistry re-equilibrates quickly, but hydrogen adjusts more slowly, leaving behind a transient shell of \htwo{} still in the process of dissociating. This excess of molecular hydrogen and corresponding dearth of atomic hydrogen drives the unique chemical signature we observe.

\item As a consequence of this chemical disequilibrium, the clouds' CO-to-\htwo{} conversion factor is elevated relative to equilibrium clouds in the Galactic Plane. Valid non-equilibrium models yield $X_{\rm{CO}} \approx 0.7-5\times 10^{21}$ cm$^{-2}$ (K km s$^{-1}$)$^{-1}$, an order of magnitude larger than the value typically assumed for disc clouds. 

\item The molecular mass of the wind is substantially larger than previous estimates. Applying our inferred $X_{\rm{CO}}$ values, we find molecular masses of $M_{\rm{mol}} \approx 1.4- 8.5 \times 10^{3}  M_\odot$ for C1 and $M_{\rm{mol}} = 2.5 - 17.6 \times 10^{3} M_\odot$ for C2, at least an order of magnitude larger than prior estimates. The molecular content of the Milky Way's nuclear wind is therefore more significant than previously thought, and the mass flux in the inner $\pm 1$ kpc of the wind is dominated by the molecular component.

\end{enumerate}

Our findings establish non-equilibrium chemistry as a central component of interpreting wind-entrained clouds and demonstrate that chemical diagnostics can provide powerful constraints on the physics of multiphase galactic winds. 

%%%%%%%%%%%%%%%%%%%%%%%%%%

\section*{Acknowledgements}
KN thanks Dr Michael Busch for discussions that led to the idea of this project. MRK acknowledges support from the Australian Research Council through Laureate Fellowship FL220100020. EDT was supported by the European Research Council (ERC) under grant agreement no. 10104075. NM-G is the recipient of an Australian Research Council Australian Laureate Fellowship (project number FL210100039) funded by the Australian Government. LA acknowledges support through
the Program ``Rita Levi Montalcini'' of the Italian MIUR. This research was supported by the Australian Government's National Collaborative Research Infrastructure Strategy (NCRIS), with access to computational resources provided by the National Computational Infrastructure through the National Computational Merit Allocation Scheme, award jh2. The MeerKAT telescope is operated by the South African Radio Astronomy Observatory, which is a facility of the National Research Foundation, an agency of the Department of Science and Innovation. CO observations were made with APEX under ESO proposal 0104.B-0106A. APEX is a collaboration between Max-Planck-Institut für Radioastronomie, the European Southern Observatory and the Onsala Space Observatory.

%%%%%%%%%%%%%%%%%%%%%%%%%%%%%%%%%%%%%%%%%%%%%%%%%%

\section*{Data Availability}
The codes to produce the results presented in this article are available via Zenodo at https://doi.org/10.5281/zenodo.18637747.



\bibliographystyle{mnras}
\bibliography{library} % if your bibtex file is called example.bib

%%%%%%%%%%%%%%%%%%%%%%%%%%%%%%%%%%%%%%%%%%%%%%%%%%

%%%%%%%%%%%%%%%%% APPENDICES %%%%%%%%%%%%%%%%%%%%%

\appendix

%%%%%%%%%%%%%%%%%%%%%%%%%%

\section{Constant volume density}
\label{app:constant}
As a comparison to the fiducial $n_{\rm H}\propto 1/r$ density profile adopted throughout the main body of this paper, we also explore a simplified model in which clouds have a constant volume density. All aspects of the modelling procedure are identical to those described in \autoref{sec:equil} and \autoref{sec:non-equ}, except that the radial density profile is replaced by a uniform value of $n_{\rm H}$ rather than the form given by \autoref{eqn:nh}. As in the main text, we focus on C1 in the R1CR1 environmental conditions.

\begin{figure}
	\includegraphics[width=\columnwidth]{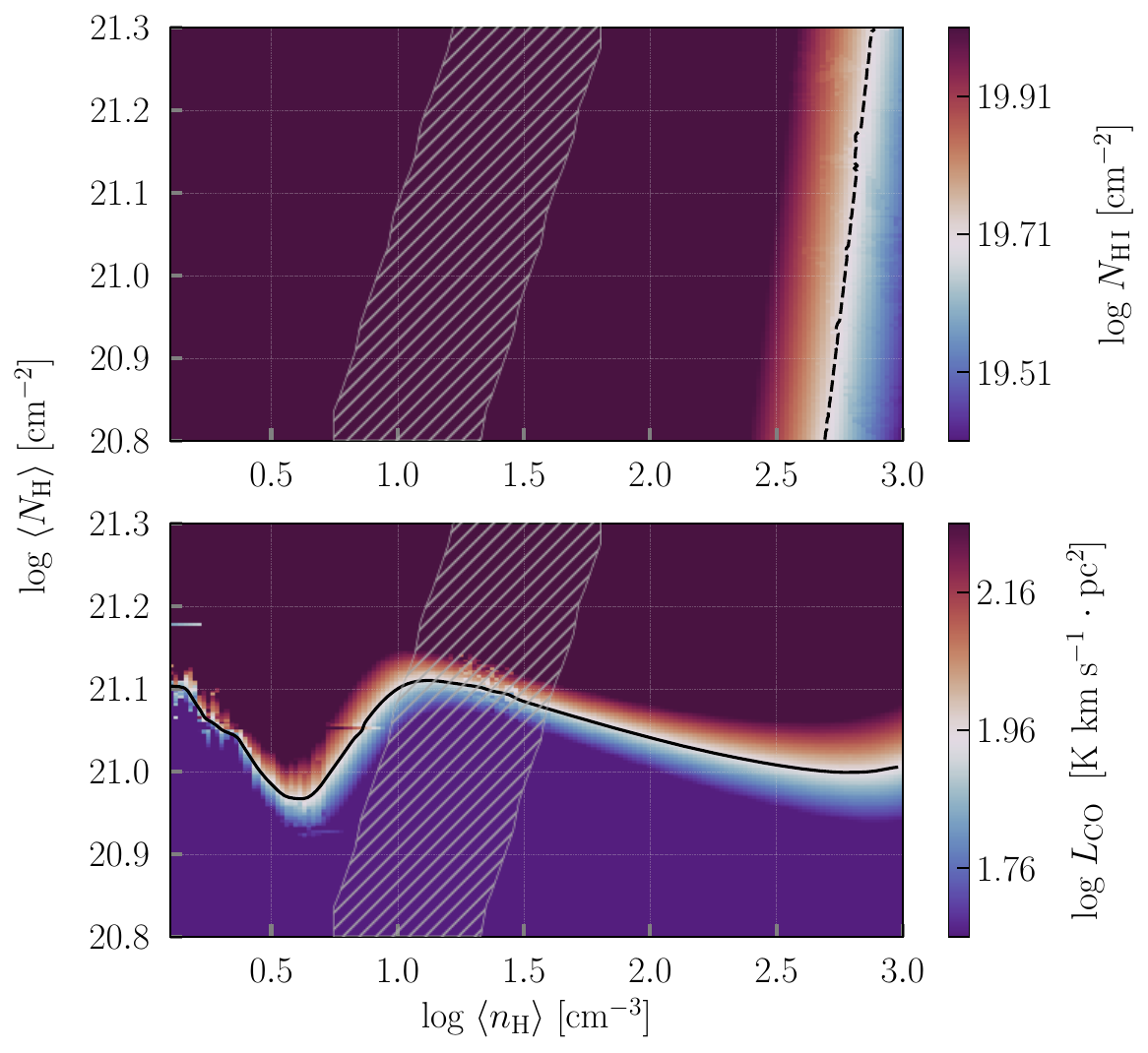}
    \caption{The same as \autoref{fig:PS1}, but for the C1 R1CR1 case assuming a constant volume density.}
    \label{fig:PS1-const}
\end{figure}

The resulting equilibrium values of \nhi{} and \lco{} as functions of \aNH{} and \anH{} are shown in \autoref{fig:PS1-const}. As in the fiducial case, there exists a region of parameter space that reproduces the observed \nhi{} and \lco{} simultaneously in chemical equilibrium. However, the corresponding parameters ($\log(\anH{}/\mathrm{cm}^{-3})=2.76$ and $\log(\aNH{}/\mathrm{cm}^{-2})=21.0$) imply a cloud radius of $R \approx 2$ pc, far smaller than the observed size. Thus, the constant-density model fails to reproduce all observed cloud properties in chemical equilibrium.

\begin{figure}
	\includegraphics[width=\columnwidth]{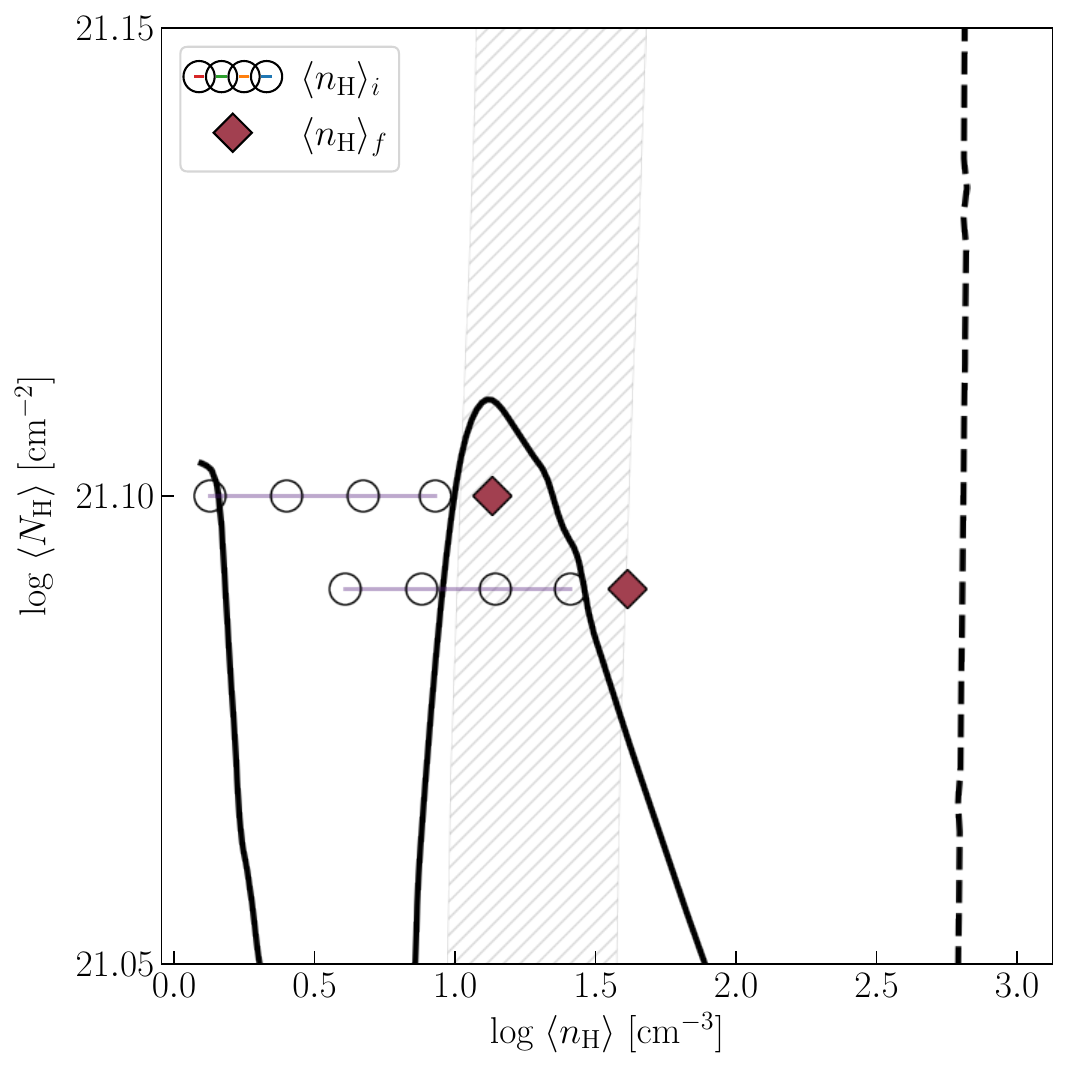}
    \caption{Same as \autoref{fig:intersection_markers}, but for the C1 R1CR1 constant-density case.}
\label{fig:intersection_markers_constant}
\end{figure}

We therefore proceed to explore non-equilibrium models, selecting initial conditions that yield clouds with physical sizes comparable to the observed cloud radius. The locations of these models in the \anH{}–\aNH{} space are shown in \autoref{fig:intersection_markers_constant}. In contrast to the $1/r$ density profile, none of the constant-density configurations pass simultaneously through the observed \nhi{} and \lco{} ranges. That is, there are no non-equilibrium constant-density models that satisfy all observational constraints at any time during its evolution.

In a $n_{\rm H}\propto 1/r$ cloud, the density peaks toward the centre, producing a compact, molecule-rich core surrounded by a lower-density \hi{}-dominated envelope. Stripping preferentially removes this diffuse envelope, rapidly reducing \nhi{} while leaving the molecular core largely intact. By contrast, a constant-density cloud lacks a dense central core. As a result, stripping primarily reduces the cloud’s physical size without substantially altering its internal chemical structure.

\begin{figure}
\includegraphics[width=\columnwidth]{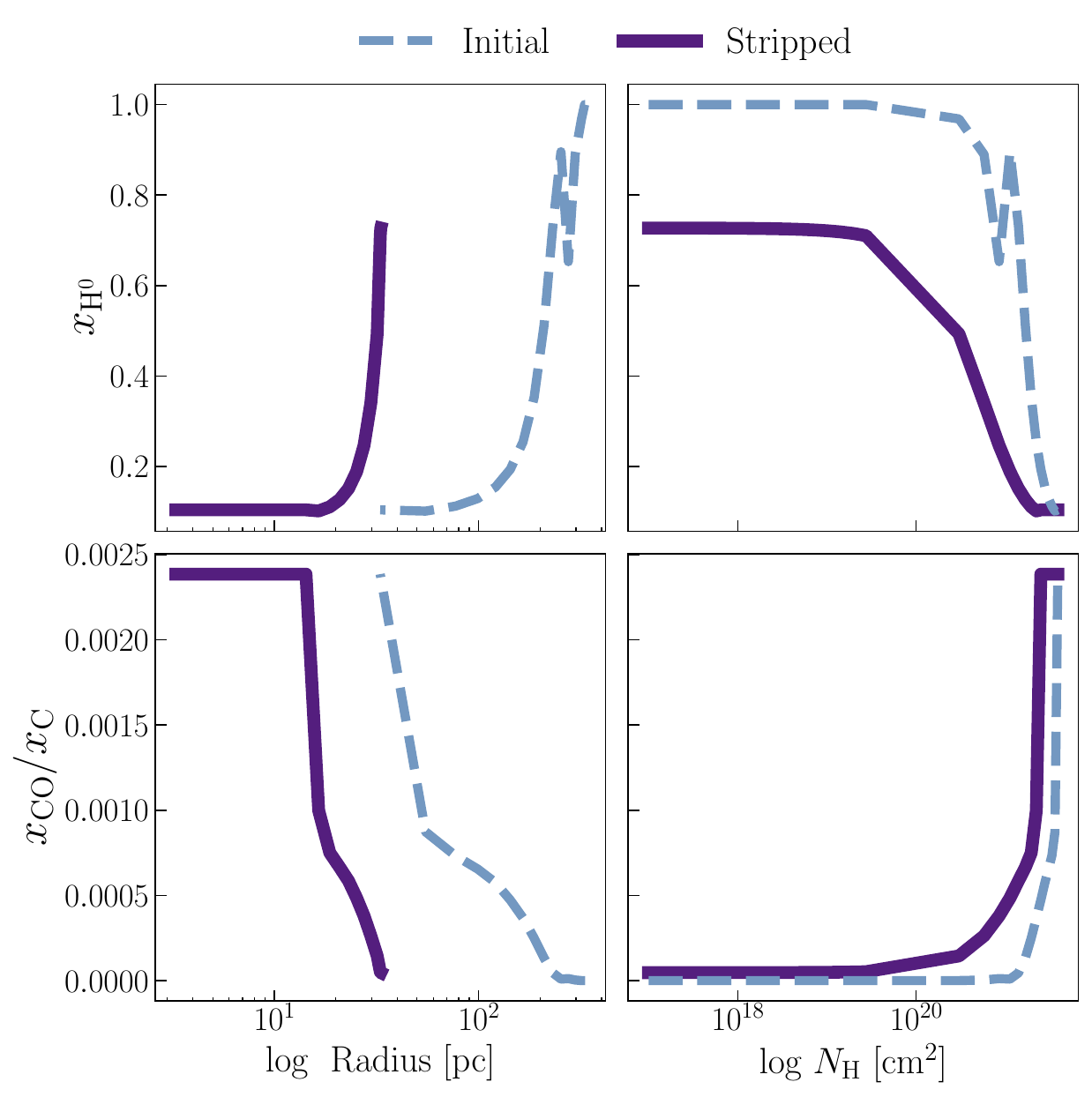}
    \caption{The same as \autoref{fig:abd_stripped}, but for the C1 R1CR1 constant-density case.}
\label{fig:abd_const}
\end{figure}

This behaviour can be seen in \autoref{fig:abd_const}, which shows that the radial abundance profiles before and after stripping retain the same shape, only differing by a truncation of the radius. As such, stripping does not drive the constant volume density cloud out of chemical equilibrium as it does the $n_{\rm H}\propto 1/r$ cloud. Further, stripping does not produce the initial reduction in \nhi{} required to match the observations. 

\begin{figure}
\includegraphics[width=\columnwidth]{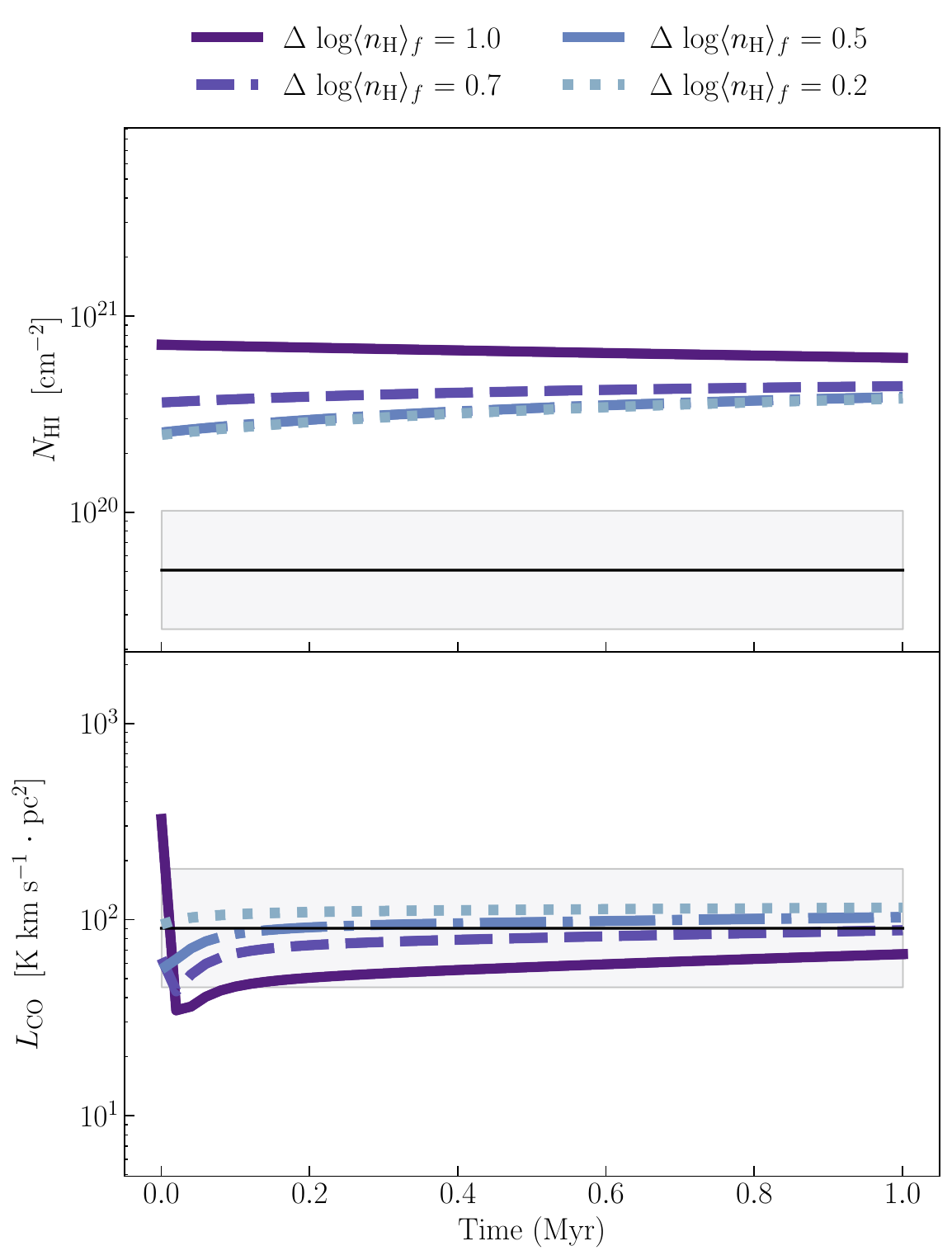}
    \caption{The same as \autoref{fig:evolve_1}, but for the C1 R1CR1 constant-density case.}
\label{fig:evo_const}
\end{figure}

As a result, the time evolution of the stripped constant-density models exhibits minimal variation in \nhi{}. The clouds begin their evolution near the equilibrium \nhi{} values, never passing through the observed region. An example of the evolution of a constant-density model is shown in \autoref{fig:evo_const} with $\log(\anH{}_i/\mathrm{cm}^{-3})=0.61$, $\log(\anH{}_f/\mathrm{cm}^{-3})=1.61$ and $\log(\aNH{}/\mathrm{cm}^{-2})=21.1$ (the rightmost diamond in \autoref{fig:intersection_markers_constant}). We therefore conclude that, in order to reproduce the observations, clouds must contain significant density contrasts.

%%%%%%%%%%%%%%%%%%%%%%%%%%

\section{Numerical convergence}
\label{app:zc}

For all our calculations in the main text we use 32 zones in our \desp{}~calculations. To test whether this is sufficient to calculate our quantities of interest, we repeat one of our representative models -- the C1, R1CR1 case with $\log(\anH{}_i/\mathrm{cm}^{-3})=0.85$, $\log(\anH{}_f/\mathrm{cm}^{-3})=1.85$ and $\log(\aNH{}/\mathrm{cm}^{-2})=21.322$ (most stripped model in row two of \autoref{tab:models}) with 64 zones instead of 32. 

\begin{figure*}
    \centering
    \begin{subfigure}[t]{0.5\textwidth}
        \centering
        \includegraphics[width=\columnwidth]{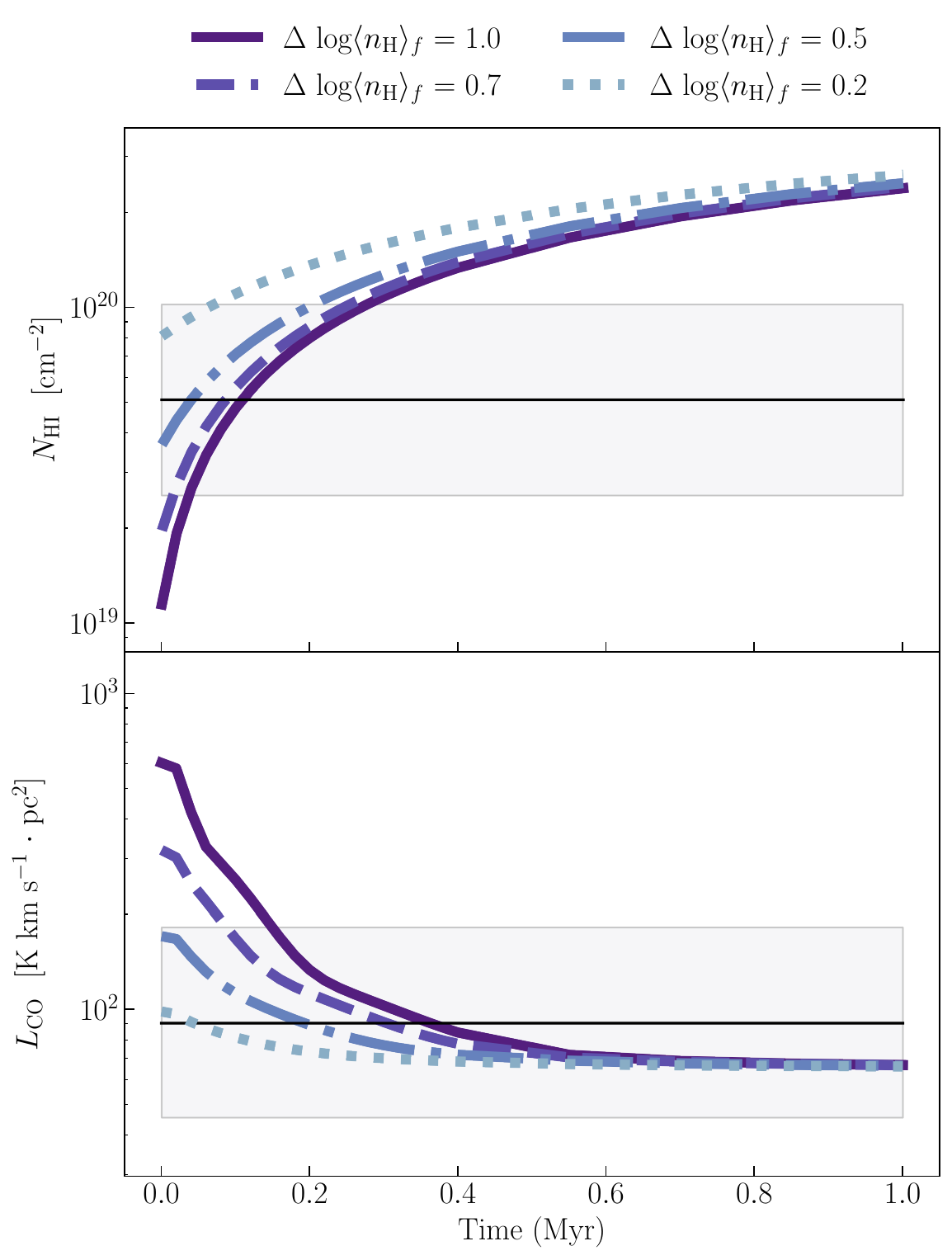}
        \caption{32 zones}
    \end{subfigure}%
    ~ 
    \begin{subfigure}[t]{0.5\textwidth}
        \centering
        \includegraphics[width=\columnwidth]{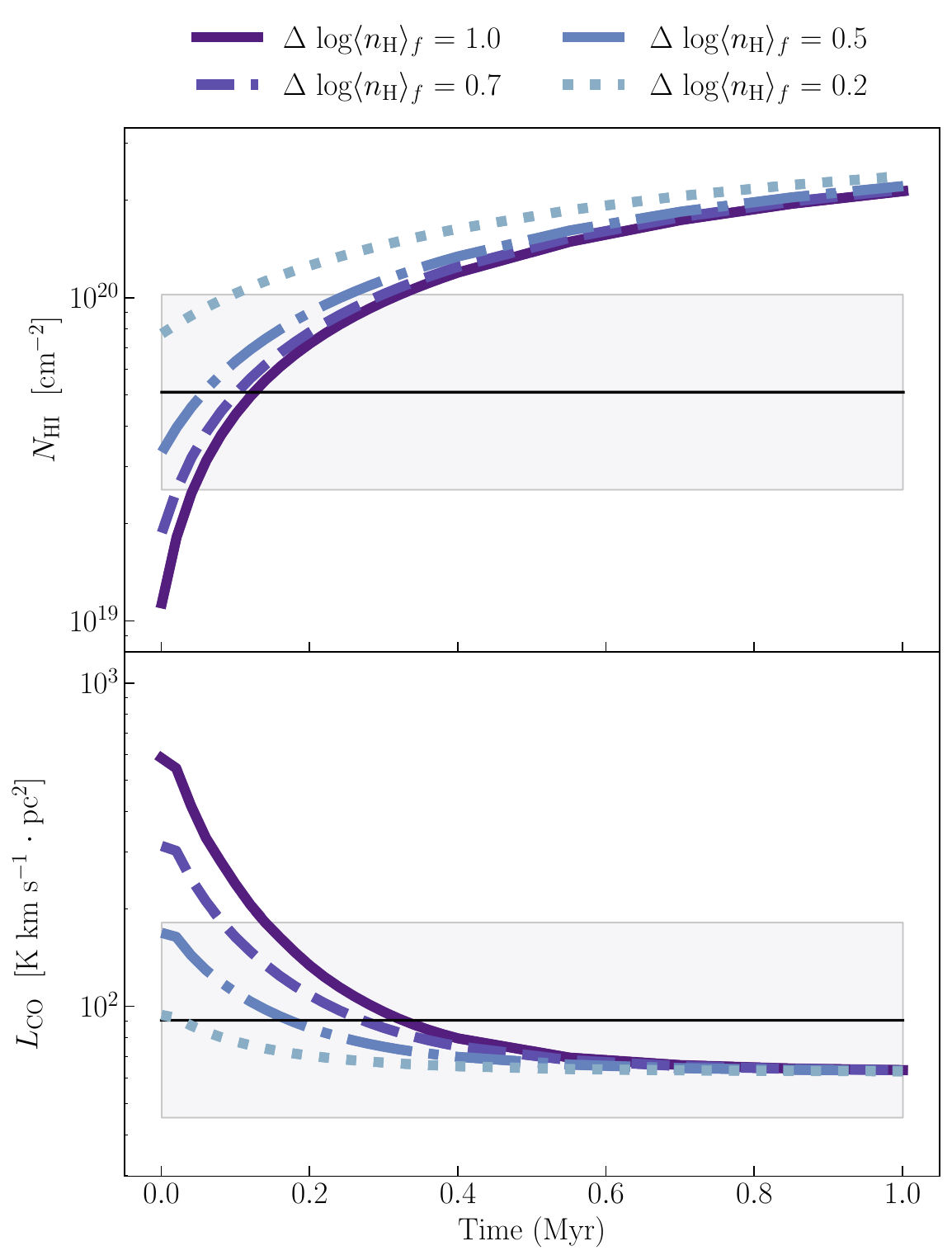}
        \caption{64 zones}
    \end{subfigure}
    
    \caption{Same as \autoref{fig:evolve_1}, but now comparing the results computed using 32 zones (left) and 64 zones (right) in \desp. The curves are near-identical, indicating that the results are well-converged.}
    \label{fig:32vs64}
\end{figure*}

We find that increasing the number of zones from 32 to 64 changes the stripped \nhi{} and \lco{} values by $1\%$ and $2\%$, respectively. The two panels in \autoref{fig:32vs64} show the time evolution of \nhi{} and \lco{} for the 32 (left) and 64 zoned (right) clouds. These small differences demonstrate that the cloud's physical and chemical properties are well converged at a resolution of 32 zones, which we therefore adopt for all models. 

%%%%%%%%%%%%%%%%%%%%%%%%%%

\section{Effects of varying the ISRF and ionisation rate}
\label{app:enviro}

In the main text we present results from the R1CR1 model for the ISRF and cosmic ray ionisation rate, and in this appendix we repeat our analysis for the other radiation environments listed in \autoref{tab:RFIR}.

\subsection{Equilibrium cloud models}

\begin{figure*}
    \centering
    \begin{subfigure}[t]{0.5\textwidth}
        \centering
        \includegraphics[width=\textwidth]{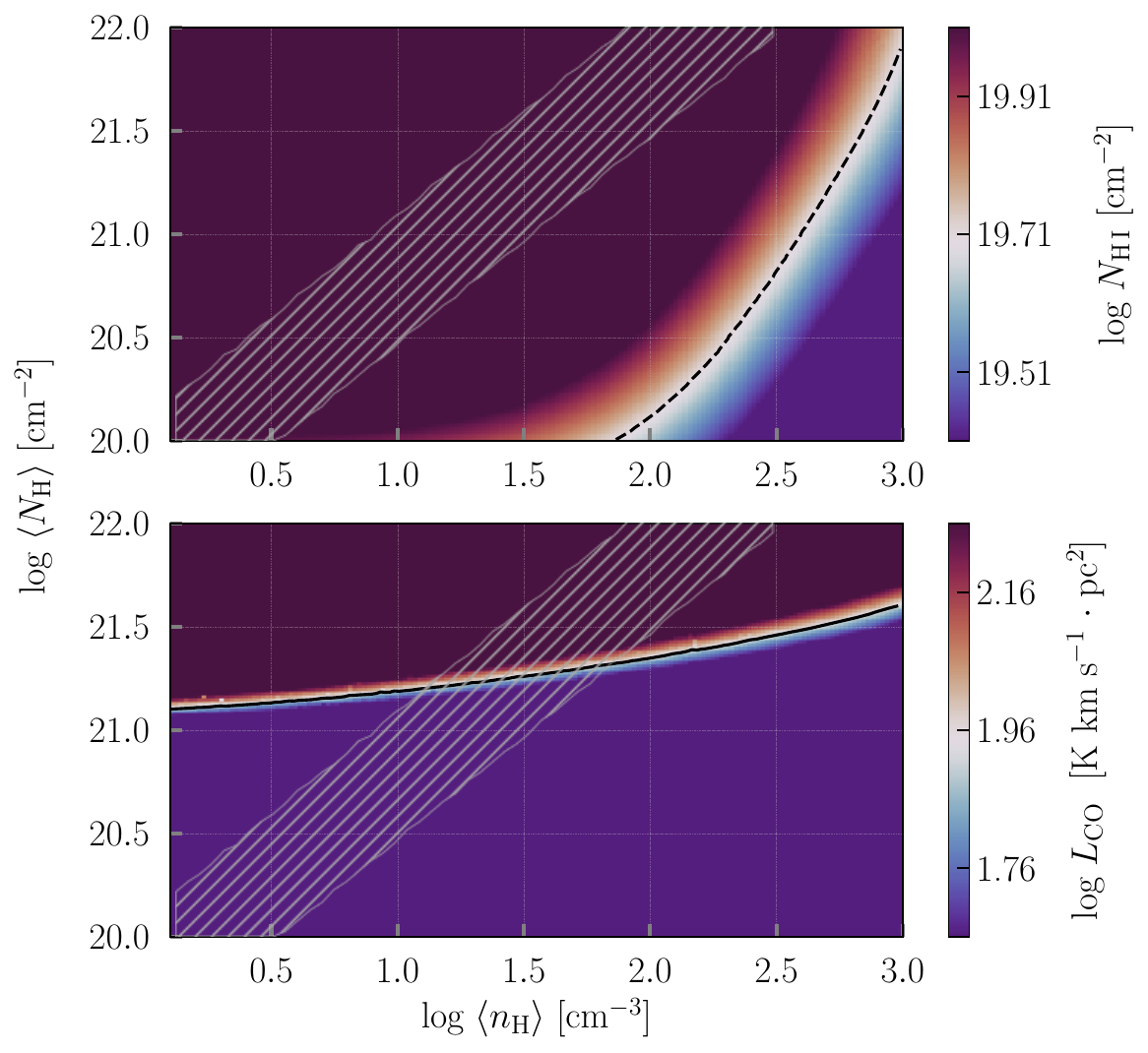}
        \caption{R1CR2}
    \end{subfigure}%
    ~ 
    \begin{subfigure}[t]{0.5\textwidth}
        \centering
        \includegraphics[width=\textwidth]{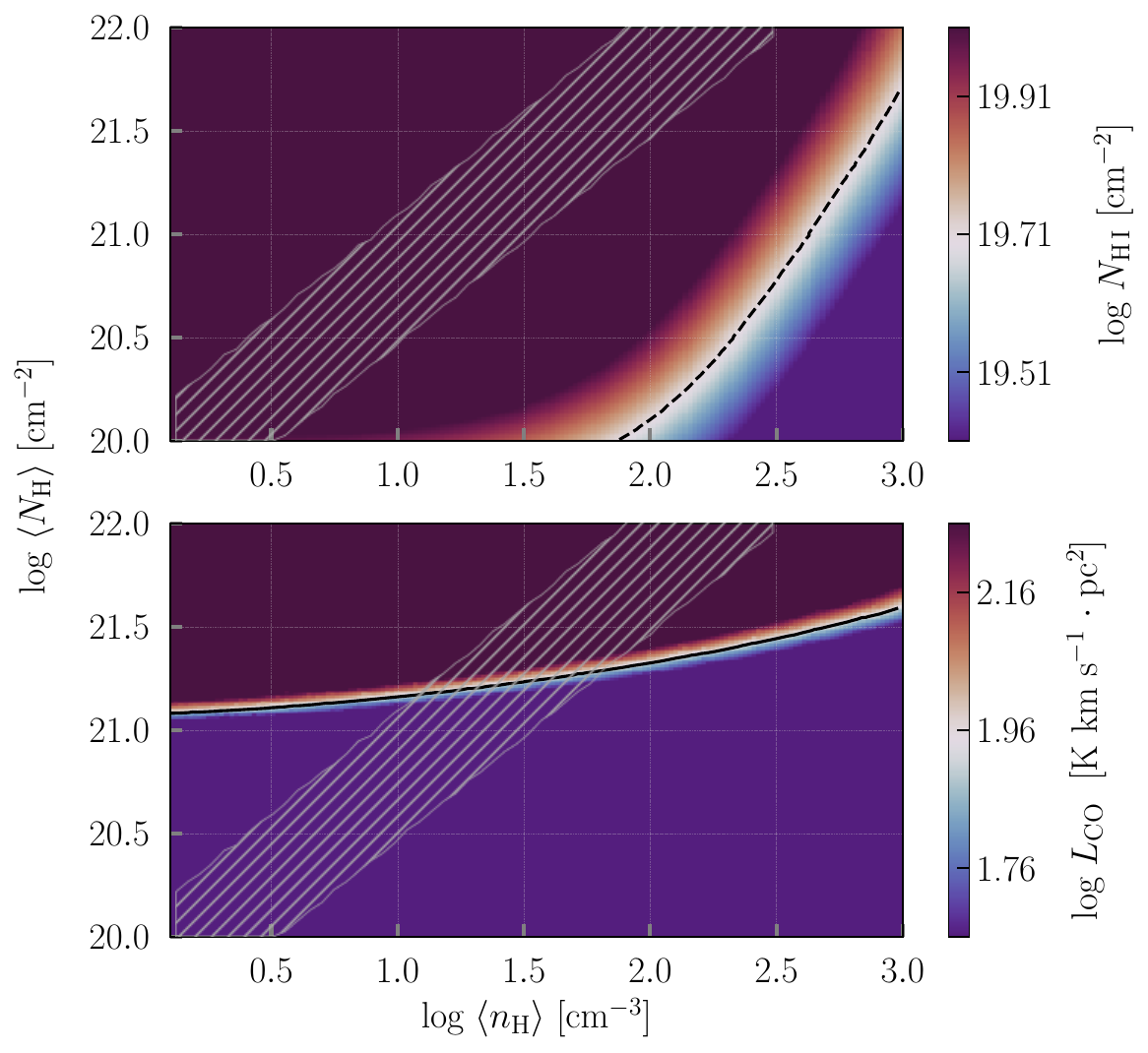}
        \caption{R1CR3}
    \end{subfigure}
    
    \begin{subfigure}[t]{0.5\textwidth}
        \centering
        \includegraphics[width=\textwidth]{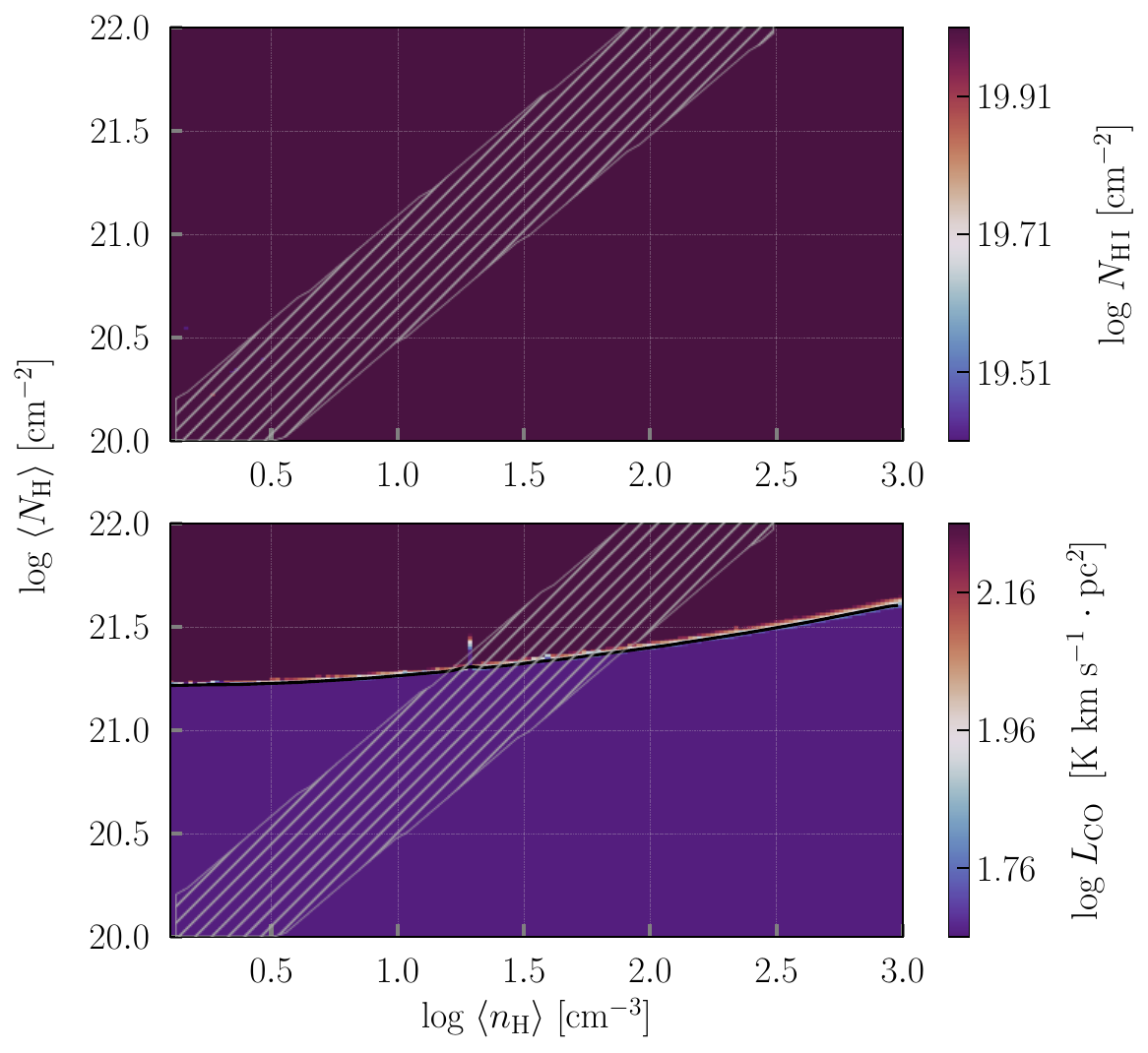}
        \caption{R2CR1}
    \end{subfigure}%
    ~ 
        \begin{subfigure}[t]{0.5\textwidth}
        \centering
        \includegraphics[width=\textwidth]{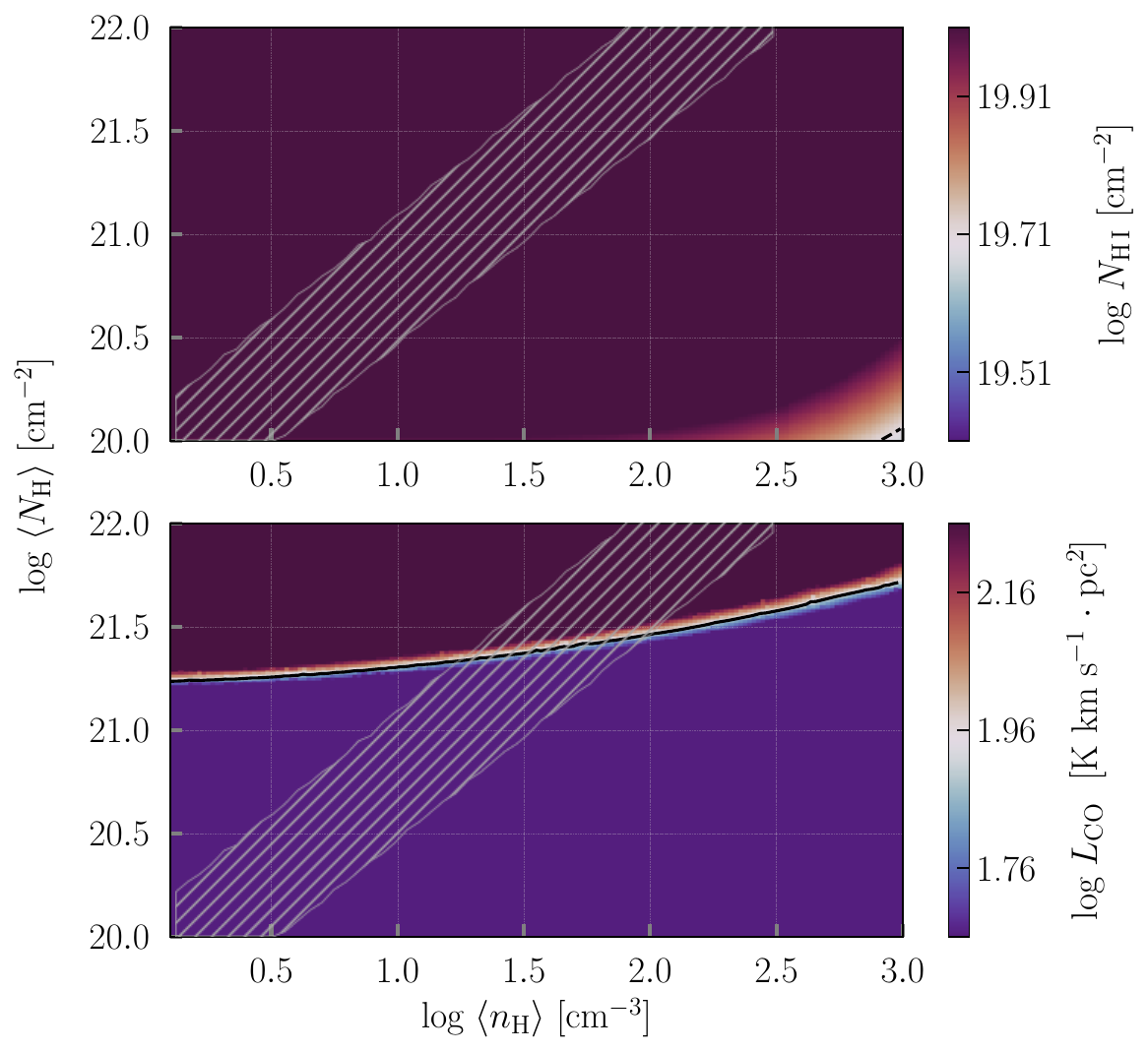}
        \caption{R2CR3}
    \end{subfigure}%
   
    \caption{Same as \autoref{fig:PS1}, but for the R1CR2, R1CR3, R2CR1 and R2CR3 radiation environments (panels a - d).}
    \label{fig:PS-all}
\end{figure*}

Repeating the equilibrium modelling presented in \autoref{sec:equil} for the remaining radiation environments yields maps of equilibrium \nhi{} and \lco{} that we show in \autoref{fig:PS-all} for cloud C1; results for C2 are nearly identical. We see that, similar to the R1CR1 scenario, the R1CR2 and R1CR3 environments produce an equilibrium solution where \nhi{} and \lco{} lie close to their observed values. However, as with R1CR1, these solutions have unrealistically-high densities and radii orders of magnitude smaller than the observed clouds. For C1, the equilibrium radii are $R_{R1CR2} = 1.22$ pc, and $R_{R1CR3} = 1.06$ pc, much smaller than the observed radius of $\sim 15$ pc. The R2CR1 and R2CR3 environments do not produce equilibrium solutions at all, as there is no configuration in which the observed \nhi{} and \lco{} values coincide, as shown in the (c) and (d) panels of \autoref{fig:PS-all}. 

\subsection{Non-equilibrium models}
More interesting than the equilibrium solutions are the non-equilibrium results obtained under the different environmental conditions. To obtain these, we repeat the procedure from \autoref{sec:non-equ}, considering models where the final, post-stripping state lies close to the locus where \lco~and cloud radius match those for the observed clouds. We list the full set of models we explore for each radiation environment in \autoref{tab:modelsall}; the subset that pass through a ``valid'' stage where they satisfy all observational constraints (\nhi{}, \lco{} and radius) are indicated by a star on the last column. As with the fiducial R1CR1 case, the R1CR2 case yields valid solutions for both clouds, albeit fewer than R1CR1, while the stronger radiation field case (R2CR1) produces valid solutions only for C1, and the strongest cosmic ray ionisation rate case (R1CR3) produces valid solutions only for C2. The strongest radiation and ionisation rate case (R2CR3) does not yield valid solutions for C1 or C2. 
\begin{table*}
    \centering
    \begin{tabular}{ccccc}
    \hline\hline
    Cloud &Case & $\log(N_\mathrm{H})$ & $\log\langle n_\mathrm{H}\rangle_f$ & $\log\langle n_\mathrm{H}\rangle_i$\\
    & & [cm$^{-2}$] & [cm$^{-3}$] & [cm$^{-3}$] \\
    \hline\hline
    
    \multirow{8}{*}{C1} & R1CR2 & 21.24 & 1.50 & 0.50$^{\bigstar}$, 0.77$^{\bigstar}$, 1.04, 1.30 \\
          && 21.29 & 1.80 & 0.80$^{\bigstar}$, 1.07$^{\bigstar}$, 1.33$^{\bigstar}$, 1.60 \\ 
    & R1CR3 & 21.17 & 1.15 & 0.15, 0.42, 0.68, 0.95 \\
          && 21.41 & 1.70 & 0.70, 0.97, 1.23, 1.50 \\
    & R2CR1 & 21.28 & 1.26 & 0.26, 0.53, 0.80, 1.06 \\
          && 21.34 & 1.79 & 0.79$^{\bigstar}$, 1.06, 1.32, 1.59 \\
    & R2CR3 & 21.32 & 1.30 & 0.30, 0.57, 0.83, 1.10 \\
          && 21.41 & 1.86 & 0.86, 1.13, 1.39, 1.66 \\
    \hline
    \multirow{8}{*}{C2} & R1CR2 & 21.14 & 1.15 & 0.15$^{\bigstar}$, 0.42, 0.68, 0.95 \\
          && 21.19 & 1.50 & 0.50$^{\bigstar}$, 0.77$^{\bigstar}$, 1.03$^{\bigstar}$, 1.30 \\
    &R1CR3 & 21.09 & 0.90 & 0.10, 0.30, 0.50, 0.70 \\
          && 21.24 & 1.52 & 0.52, 0.78$^{\bigstar}$, 1.05$^{\bigstar}$, 1.32 \\
    &R2CR1 & 21.36 & 1.21 & 0.21, 0.48, 0.74, 1.01 \\
          && 21.24 & 1.79 & 0.79, 1.06, 1.32, 1.59 \\
    &R2CR3 & 21.27 & 1.10 & 0.10, 0.37, 0.64, 0.90 \\
          && 21.35 & 1.68 & 0.68, 0.95, 1.22, 1.48 \\
    \hline
    $^{\bigstar}$ Valid model.
    \end{tabular}
    \caption{Parameters of non-equilibrium chemical models explored for the R1CR2, R1CR3, R2CR1 and R2CR3 cases. Each model is characterised by a mean cloud column density \aNH, a mean volume density before stripping \anH$_i$, and a mean volume density after stripping \anH$_f$. }
    \label{tab:modelsall}
\end{table*}

 \begin{table}
    \centering
    \begin{tabular}{cccc}
    \hline\hline
    Cloud & Case & $X_\mathrm{CO(2\to 1)}$ & $X_\mathrm{CO(1\to 0)}$ \\
    
    & & (cm$^{-2}$) (K km s$^{-1}$)$^{-1}$ & (cm$^{-2}$) (K km s$^{-1}$)$^{-1}$ \\
    \hline\hline
    
    \multirow{2}{*}{C1} & R1CR2 & $1.15-3.14\times10^{21}$ & $0.81-2.08\times10^{21}$ \\ 
    & R2CR1 & $3.66\times10^{21}$ & $4.14\times10^{21}$ \\ 
    \hline

    \multirow{2}{*}{C2} & R1CR2 & $1.53-5.34\times10^{21}$ & $1.10-3.56\times10^{21}$ \\ 
    & R1CR3 & $1.38\times10^{21}$ & $1.05\times10^{21}$ \\ 
    \hline
    \end{tabular}
    \caption{The range of $X_\mathrm{CO(2\to 1)}$ and $X_\mathrm{CO(1\to 0)}$ values for valid non-equilibrium models for C1 and C2 across the various cases explored.}
    \label{tab:xcoall}
\end{table}

We can understand the general pattern that we find fewer and fewer valid models as we increase the radiation field and cosmic ionisation rate as follows. Stronger radiation fields accelerate photodissociation, while higher cosmic ray ionisation rates drive the H$^0$-\htwo{} transition zone to higher densities deeper within the cloud. Both effects reduce chemical equilibration timescales -- the former by supplying H$_2$-dissociating photons at a larger rate, and the latter by pushing the region where the atomic-molecular transition occurs to denser gas where collision rates are higher. In either case, the non-equilibrium states required to match the low observed \nhi{} become increasingly short-lived and therefore less likely to be observed, progressively restricting the viable parameter space. This finding suggests that the radiation environment in which the observed clouds reside cannot be too intense, but it is hard to make a more quantitative statement without a much denser exploration of parameter space than we have performed.

Nonetheless, when valid solutions do exist, the resulting CO conversion factors, $X_\mathrm{CO}$, remain similar to those found for R1CR1. We report these values in \autoref{tab:xcoall}. This indicates that our conclusions regarding $X_\mathrm{CO}$ are insensitive to environmental conditions within the range that permits valid non-equilibrium states.

%%%%%%%%%%%%%%%%%%%%%%%%%%%%%%%%%%%%%%%%%%%%%%%%%%

% Don't change these lines
\bsp	% typesetting comment
\label{lastpage}
\end{document}